\documentclass[reprint, lengthcheck, aps, prd, nofootinbib, floatfix,%
amsmath, amssymb, amsfonts, showpacs, superscriptaddress]{revtex4}

\pdfoutput=1

\usepackage{graphicx} %
\usepackage{color} %
\usepackage{bm} %

\usepackage[all]{onlyamsmath} % Error on bad non-ams math

\usepackage{braket}
\usepackage{accents} %
\usepackage{xspace} % Adds a space after a text macro
\input{Preamble}
\renewcommand{\d}{\ensuremath{\mathrm{d}}}
\newcommand{\e}{\ensuremath{\mathrm{e}}}
\renewcommand{\i}{\ensuremath{\mathrm{i}}}

\newcommand{\Mirr}{\ensuremath{M_{\text{irr}}}}
\newcommand{\Eadm}{\ensuremath{M_{\text{ADM}}}}

\newcommand{\ADMMass}{\Eadm}
\newcommand{\IrrMass}{\Mirr}

\newcommand{\tr}{\ensuremath{t_{\text{ret}}}}
\newcommand{\trs}{\ensuremath{\accentset{\smile}{t}_{\text{ret}}}}
\newcommand{\tc}{\ensuremath{T}}
\newcommand{\tcorr}{\ensuremath{t_{\text{corr}}}}
\newcommand{\rc}{\ensuremath{R}}
\newcommand{\ra}{\ensuremath{R_{\text{areal}}}}
\newcommand{\rt}{\ensuremath{r_{\ast}}}

\newcommand{\h}{\ensuremath{h}}

\newcommand{\Sun}{\ensuremath{{M}_\odot}}
\newcommand{\MSun}{\Sun}

\newcommand{\foreign}[1]{\textrm{#1}}
\newcommand{\etal}{\foreign{et~al.}\xspace}
\newcommand{\eg}{\foreign{e.g.}\xspace}
\newcommand{\ie}{\foreign{i.e.}\xspace}

\definecolor{NoteColor}{rgb}{0,0,.85}

\definecolor{NewColor}{rgb}{0,.55,0}

\newcommand{\software}[1]{\textsc{#1}}
\newcommand{\command}[1]{\texttt{#1}}

\newcommand{\CapName}[1]{\textbf{#1}.  }

\makeatletter
\let\protect\relax
{\catcode`\|=\active
  \xdef\InnerProduct{\protect\expandafter\noexpand\csname InnerProduct \endcsname}
  \expandafter\gdef\csname InnerProduct \endcsname#1{%
    \begingroup
    \ifx\SavedDoubleVert\relax
    \let\SavedDoubleVert\|\let\|\IpDoubleVert
    \fi
    \mathcode`\|32768\let|\IPVert
    \left({#1}\right)
    \endgroup
  }
}
\def\IPVert{\@ifnextchar|{\|\@gobble}% turn || into \|
     {\egroup\,\mid@vertical\,\bgroup}}
\def\IPDoubleVert{\egroup\,\mid@dblvertical\,\bgroup}
\let\SavedDoubleVert\relax
\def\midvert{\egroup\mid\bgroup}
\def\SetVert{\@ifnextchar|{\|\@gobble}% turn || into \|
    {\egroup\;\mid@vertical\;\bgroup}}
\def\SetDoubleVert{\egroup\;\mid@dblvertical\;\bgroup}
\def\mid@vertical{\mskip1mu\vrule\mskip1mu}
\def\mid@dblvertical{\mskip1mu\vrule\mskip2.5mu\vrule\mskip1mu}
\makeatother
\newcommand{\Overlap}{\Braket}

\newcommand{\Caltech}{\affiliation{Theoretical Astrophysics 350-17,
    California Institute of Technology, Pasadena, CA 91125}} %

\newcommand{\Cornell}{\affiliation{Center for Radiophysics and
    Space Research, Cornell University, Ithaca, New York, 14853}} %

\newcommand{\CITA}{\affiliation{Canadian Institute for Theoretical
    Astrophysics, 60 St.~George Street, University of Toronto,
    Toronto, ON M5S 3H8, Canada}} %

\begin{document}

%%%%%%%%%%%%%%%%

\graphicspath{{Plots/}}

\title{Extrapolating gravitational-wave data from numerical
  simulations}

\author{Michael Boyle} \Cornell \Caltech %
\author{Abdul H. Mrou\'e} \Cornell \CITA %

\date{\today}

\begin{abstract}
  Two complementary techniques are developed for obtaining the
  asymptotic form of gravitational-wave data at large radii from
  numerical simulations, in the form of easily implemented algorithms.
  It is shown that, without extrapolation, near-field effects produce
  errors in extracted waveforms that can significantly affect LIGO
  data analysis.  The extrapolation techniques are discussed in the
  context of Newman--Penrose data applied to extrapolation of
  waveforms from an equal-mass, nonspinning black-hole binary
  simulation.  The results of the two methods are shown to agree
  within error estimates.  The various benefits and deficiencies of
  the methods are discussed.
\end{abstract}

% 04.25.D- Numerical relativity %
% 04.25.dg Numerical studies of black holes and black-hole binaries %
% 04.25.Nx Post-Newtonian approximation; perturbation theory; related
% approximations %
% 04.30.-w Gravitational waves (see also 04.80.Nn: Gravitational-wave
% detectors and experiments) %
% 04.30.Db Wave generation and sources %
% 02.70.Hm Spectral methods

\maketitle

%%%%%%%%%%%%%%%%%%%%%%%%%%%%%%%%%%%%%%%%%%%%%%%%%%%%%%%%%%%%%%%%%%%%%%
%%%%%%%%%%%%%%%%%%%%%%%%%%%%%%%%%%%%%%%%%%%%%%%%%%%%%%%%%%%%%%%%%%%%%%
\section{Introduction}
\label{Sec:Introduction}

As numerical simulations of black-hole binaries improve, the criterion
for success moves past the ability of a code to merely persist through
many orbits of inspiral, merger, and ringdown.  Accuracy becomes the
goal, as related work in astrophysics and analysis of data from
gravitational-wave detectors begins to rely more heavily on results
from numerical relativity.  One of the most important challenges in
the field today is to find and eliminate systematic errors that could
pollute results built on numerics.  Though there are many possible
sources of such error, one stands out as being particularly easy to
manage and---as we show---a particularly large effect: the error made
by extracting gravitational waveforms from a simulation at finite
radius, and treating these waveforms as though they were the
asymptotic form of the radiation.

The desired waveform is the one to which post-Newtonian approximations
aspire, and the one sought by gravitational-wave observatories: the
asymptotic waveform.  This is the waveform as it is at distances of
over $10^{14}\,M$ from the system generating the waves.  In typical
numerical simulations, data extraction takes place at a distance of
order $100\,M$ from the black holes.  At this radius, the waves are
still rapidly changing because of real physical effects.  Near-field
effects~\cite{Teukolsky1982, Boyle2008, Boyle2009} are plainly
evident, scaling with powers of the ratio of the reduced wavelength to
the radius, $(\lambdabar/r)^k$.\footnote{We use the standard notation
  $\lambdabar \equiv \lambda / 2\pi$.} %
Extraction methods aiming to eliminate the influence of gauge effects
alone (\eg, improved Regge--Wheeler--Zerilli or quasi-Kinnersley
techniques) will not be able to account for these physical changes.

Even using a rather naive, gauge-dependent extraction method,
near-field effects dominate the error in extracted waves throughout
the inspiral for the data presented in this paper~\cite{Boyle2008}.
For extraction at $r=50\,M$, these effects can account for a
cumulative error of roughly $50\%$ in amplitude or a phase difference
of more than one radian, from beginning to end of a 16-orbit
equal-mass binary merger.  Note that near-field effects should be
proportional to---at leading order---the ratio of $\lambdabar/r$ in
phase and $(\lambdabar/r)^2$ in amplitude, as has been observed
previously \cite{HannamEtAl2008, Boyle2008}.  Crucially, because the
wavelength changes most rapidly during the merger, the amplitude and
phase differences due to near-field effects also change most rapidly
during merger.  This means that coherence is lost between the inspiral
and merger/ringdown segments of the waveform.

We can see the importance of this decoherence by looking at its effect
on the matched-filtering technique frequently used to analyze data
from gravitational-wave detectors.  Matched filtering~\cite{Finn1992,
  FinnChernoff1993, BoyleEtAl2009a} compares two signals, $s_{1}(t)$
and $s_{2}(t)$.  It does this by Fourier transforming each into the
frequency domain, taking the product of the signals, weighting each
inversely by the noise---which is a function of frequency---and
integrating over all frequencies.  This match is optimized over the
time and phase offsets of the input waveforms.  For appropriately
normalized waveforms, the result is a number between 0 and 1, denoted
$\Overlap{ s_{1} | s_{2}}$, with 0 representing no match, and 1
representing a perfect match.  If we take the extrapolated waveform as
$s_{1}$ and the waveform extracted at finite radius as $s_{2}$, we can
evaluate the match between them.  If the extrapolated waveform
accurately represents the ``true'' physical waveform, the mismatch
(defined as $1-\Overlap{s_{1}|s_{2}}$) shows us the loss of signal in
data analysis if we were to use the finite-radius waveforms to search
for physical waveforms in detector data.

\begin{figure}
  %MatchInitial
  \includegraphics[width=\linewidth]{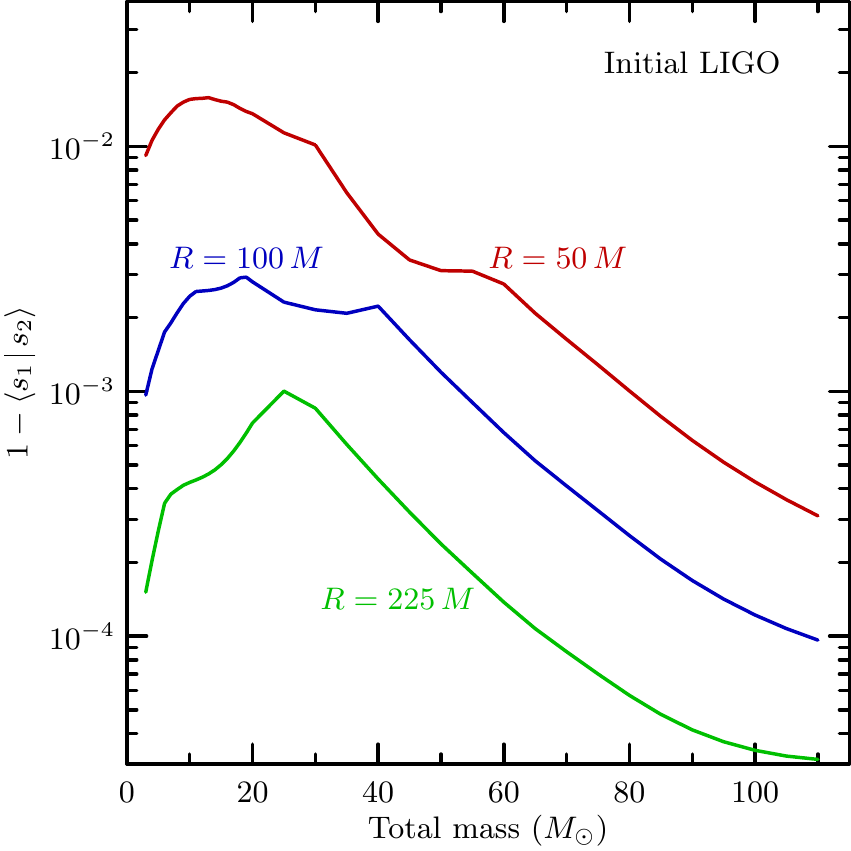}
  \caption{\CapName{Data-analysis mismatch between finite-radius
      waveforms and the extrapolated waveform for Initial LIGO} This
    plot shows the mismatch between extrapolated waveforms and
    waveforms extracted at several finite radii, scaled to various
    values of the total mass of the binary system, using the
    Initial-LIGO noise curve.  The waveforms are shifted in time and
    phase to find the optimal match.  Note that the data used here is
    solely numerical, with no direct post-Newtonian contribution.
    Thus, for masses below $40\,\MSun$, this data represents only a
    portion of the physical waveform.}
  \label{fig:MismatchInitial}
\end{figure}

\begin{figure}
  %MatchAdvanced
  \includegraphics[width=\linewidth]{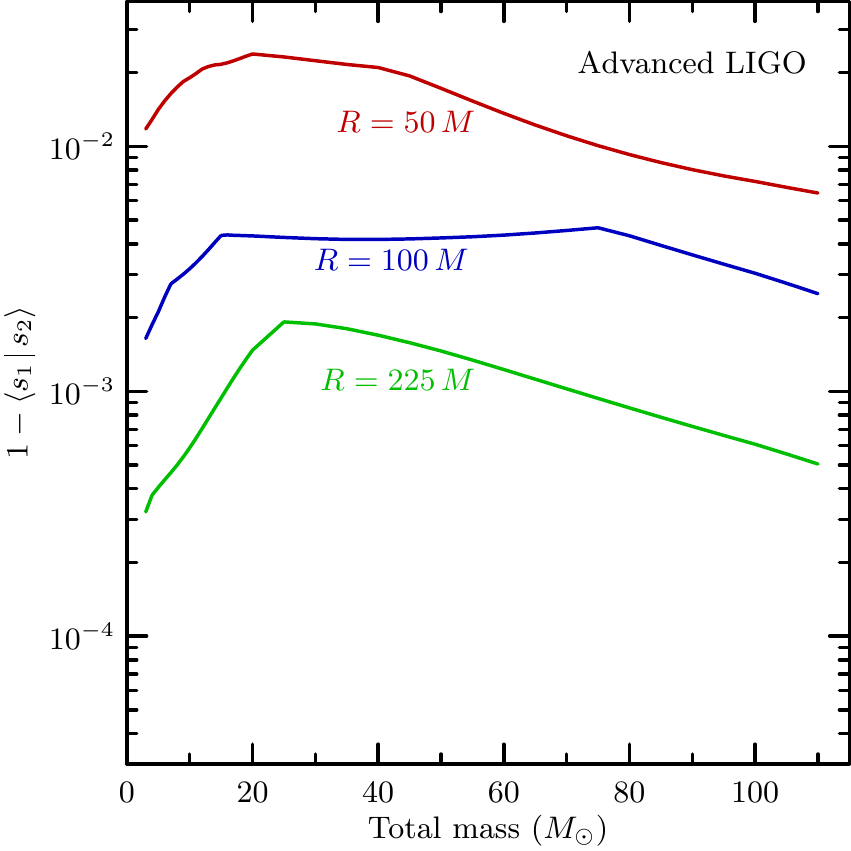}
  \caption{\CapName{Data-analysis mismatch between finite-radius
      waveforms and the extrapolated waveform for Advanced LIGO} This
    plot shows the mismatch between extrapolated waveforms and
    waveforms extracted at several finite radii, scaled to various
    values of the total mass of the binary system, using the
    Advanced-LIGO noise curve.  The waveforms are shifted in time and
    phase to find the optimal match.  Note that the data used here is
    solely numerical, with no direct post-Newtonian contribution.
    Thus, for masses below $110\,\MSun$, this data represents only a
    portion of the physical waveform.}
  \label{fig:MismatchAdvanced}
\end{figure}

The waveforms have a simple scaling with the total mass of the system,
which sets their frequency scale relative to the noise present in the
detector.  In Figs.~\ref{fig:MismatchInitial}
and~\ref{fig:MismatchAdvanced}, we show mismatches between
finite-radius and extrapolated data from the Caltech--Cornell group
for a range of masses of interest to LIGO data analysis, using the
Initial- and Advanced-LIGO noise curves, respectively, to weight the
matches.  The value of $R$ denotes the coordinate radius of extraction
for the finite-radius waveform.

%%%%%%%%%%%%%%%%%%%%%%%%%%%%%%%%%%%%%%%%%%%%%%%%%%%%%%%%%%%%%%%%%%%%%%
%%%%%%%%%%%%%%%%%%%%%%%%%%%%%%%%%%%%%%%%%%%%%%%%%%%%%%%%%%%%%%%%%%%%%%

%%%%%%%%%%%%%%%%%%%%%%%%%%%%%%%%%%%%%%%%%%%%%%%%%%%%%%%%%%%%%%%%%%%%%%
%%%%%%%%%%%%%%%%%%%%%%%%%%%%%%%%%%%%%%%%%%%%%%%%%%%%%%%%%%%%%%%%%%%%%%

The data in these figures is exclusively numerical data from the
simulation used throughout this paper, with no direct contributions
from post-Newtonian (PN) waveforms.  However, to reduce ``turn-on''
artifacts in the Fourier transforms, we have simply attached PN
waveforms to the earliest parts of the time-domain waveforms,
performed the Fourier transform, and set to zero all data at
frequencies for which PN data are used.  The match integrals are
performed over the intersection of the frequencies present in each
waveform, as in Ref.~\cite{HannamEtAl2009a}.

This means that the data used here is not truly complete for masses
below $40\,\MSun$ in Initial LIGO and $110\,\MSun$ in Advanced LIGO,
and that a detected signal would actually be dominated by data at
lower frequencies than are present in this data for masses below about
$10\,\MSun$.  These masses are correspondingly larger for shorter
waveforms, which begin at higher frequencies.  It is important to
remember that this type of comparison can only show that a given
waveform (of a given length) is as good as it needs to be for a
detector.  If the waveform does not cover the sensitive band of the
detector, the detection signal-to-noise ratio would presumably improve
given a comparably accurate waveform of greater duration.  Thus, the
bar is raised for longer waveforms, and for lower masses.

These figures demonstrate that the mismatch can be of order $1\%$ when
extracting at a radius of $R=50\,M$.  For extraction at $R=225\,M$,
the mismatch is never more than about $0.2\%$.  The loss in event rate
would be---assuming homogeneous distribution of events in
space---roughly 3 times the mismatch when using a template bank based
on imperfect waveforms~\cite{Brown2004}.  Lindblom
\etal~\cite{LindblomEtAl2008} cite a target mismatch of less than
$0.5\%$ between the physical waveform and a class of model templates
to be used for \emph{detection} of events in current LIGO detector
data.\footnote{ This number of 0.5\% results from assumptions about
  typical event magnitude, template bank parameters, and requirements
  on the maximum frequency of missed events.  The parameters used to
  arrive at this number are typical for Initial LIGO.}  Thus, for
example, if these numerical waveforms were to be used in construction
of template banks,\footnote{ We emphasize that these waveforms do not
  cover the sensitive band of current detectors, and thus would not
  likely be used to construct template banks without the aid of
  post-Newtonian extensions of the data.  Longer templates effectively
  have more stringent accuracy requirements, so the suitability of
  these extraction radii would change for waveforms of different
  lengths.  In particular, our results are consistent with those of
  Ref.~\cite{HannamEtAl2009a}, which included non-extrapolated data of
  shorter duration.} the waveform extracted at $R=50\,M$ would not be
entirely sufficient, in the sense that a template bank built on
waveforms with this level of inaccuracy would lead to an unacceptably
high reduction of event rate.  The waveforms extracted at $R=100\,M$
and $225\,M$, on the other hand, may be acceptable for Initial LIGO.
For the loudest signals expected to be seen by Advanced LIGO, the
required mismatch may be roughly $10^{-4}$~\cite{LindblomEtAl2008}.
In this case, even extraction at $R=225\,M$ would be insufficient;
some method must be used to obtain the asymptotic waveform.  For both
Initial and Advanced LIGO, estimating the parameters of the
waveform---masses and spins of the black holes, for
instance---requires still greater accuracy.

% \Note{We have $f_{\text{initial}} = \unit[1100 \pm 20]{Hz}
%   \frac{\Sun}{M}$ and $f_{\text{ringdown}} = \unit[17800]{Hz}
%   \frac{\Sun}{M}$.  The minimum of the LIGO PSD occurs at about
%   $\unit[153]{Hz}$, so the relevant mass range is (roughly) $8 \leq
%   M/\Sun \leq 132$.  Because of the seismic wall, the absolute
%   maximum mass of interest for an equal-mass system is $M \approx
%   400\,\Sun$.  Similarly, because of the Nyquist frequency, the
%   absolute minimum \emph{for this waveform} is $M \approx
%   0.5\,\Sun$, though that's not very relevant.}

% \Note{Evaluate some of the errors that could be present in radiated
%   energy or linear or angular momentum, if extrapolation is not
%   done.  Note that our estimates are only for equal-mass,
%   non-spinning binaries; the errors could be much larger for more
%   complicated systems, with more complicated gauge issues.}

Extrapolation of certain quantities has been used for some time in
numerical relativity.  Even papers announcing the first successful
black-hole binary evolutions \cite{BuonannoEtAl2007,
  CampanelliEtAl2006, BakerEtAl2006} showed radial extrapolation of
scalar physical quantities---radiated energy and angular momentum.
But waveforms reported in the literature have not always been
extrapolated.  For certain purposes, this is
acceptable---extrapolation simply removes one of many errors.  If the
precision required for a given purpose allows it, extrapolation is
unnecessary.  However, for the purposes of LIGO data analysis, we see
that extrapolation of the waveform may be very important.

We can identify three main obstacles to obtaining the asymptotic form
of gravitational-wave data from numerical simulations:
\begin{enumerate}
 \item Getting the ``right'' data at any given point, independent of
  gauge effects (\eg, using quasi-Kinnersley techniques and improved
  Regge--Wheeler--Zerilli techniques);
 \item Removing near-field effects;
 \item Extracting data along a physically relevant path.
\end{enumerate}
Many groups have attempted to deal with the first of these
problems.\footnote{See~\cite{NerozziEtAl2005}
  and~\cite{SarbachTiglio2001} and references therein for descriptions
  of quasi-Kinnersley and RWZ methods, respectively.  Also, an
  interesting discussion of RWZ methods, and the possibility of
  finding the ``exact'' waveform at finite distances is found
  in~\cite{PazosEtAl2007}.}  While this is, no doubt, an important
objective, even the best extraction technique to date is imperfect at
finite radii.  Moreover, at finite distances from the source,
gravitational waves continue to undergo real physical changes as they
move away from the system~\cite{Thorne1980}, which are frequently
ignored in the literature.  Some extraction techniques have been
introduced that attempt to incorporate corrections for these physical
near-field effects~\cite{AbrahamsEvans1988, AbrahamsEvans1990,
  DeadmanStewart2009}.  However, these require assumptions about the
form of those corrections, which we prefer not to impose.  Finally,
even if we have the optimal data at each point in our spacetime, it is
easy to see that extraction along an arbitrary (timelike) path through
that spacetime could produce a nearly arbitrary waveform, bearing no
resemblance to a waveform that could be observed in a nearly inertial
detector.  In particular, if our extraction point is chosen at a
specific coordinate location, gauge effects could make that extraction
point correspond to a physical path which would not represent any real
detector's motion.  It is not clear how to estimate the uncertainty
this effect would introduce to the waveforms, except by removing the
effect entirely.

We propose a simple method using existing data-extraction techniques
which should be able to overcome each of these three obstacles, given
certain very basic assumptions.  The data are to be extracted at a
series of radii---either on a series of concentric spheres, or at
various radii along an outgoing null ray.  These data can then be
expressed as functions of extraction radius and retarded time using
either of two simple methods we describe.  For each value of retarded
time, the waveforms can then be fit to a polynomial in inverse powers
of the extraction radius.  The asymptotic waveform is simply the first
nonzero term in the polynomial.  Though this method also incorporates
certain assumptions, they amount to assuming that the data behave as
radially propagating waves, and that the metric itself is
asymptotically Minkowski in the coordinates chosen for the simulation.

Extrapolation is, by its very nature, a dangerous procedure.  The
final result may be numerically unstable, in the sense that it will
fail to converge as the order of the extrapolating polynomial is
increased.  This is to be expected, as the size of the effects to be
removed eventually falls below the size of noise in the waveform data.
There are likely better methods of determining the asymptotic form of
gravitational waves produced by numerical simulations.  For example,
characteristic evolution is a promising technique that may become
common in the near future~\cite{BishopEtAl1996, Huebner2001,
  BabiucEtAl2005, BabiucEtAl2008}.  Nonetheless, extrapolation does
provide a rough and ready technique which can easily be implemented by
numerical-relativity groups using existing frameworks.

This paper presents a simple method for implementing the extrapolation
of gravitational-wave data from numerical simulations, and the
motivation for doing so.  In Sec.~\ref{sec:TortoiseExtrapolation}, we
begin by introducing an extrapolation method that uses approximate
tortoise coordinates, which is the basic method used to extrapolate
data in various papers~\cite{BoyleEtAl2007b, BoyleEtAl2008a,
  ScheelEtAl2009, BoyleEtAl2009a, BuonannoEtAl2009a} by the
Caltech--Cornell collaboration.  The method is tested on the inspiral,
merger, and ringdown waveform data of the equal mass, nonspinning,
quasicircular 15-orbit binary simulation of the Caltech--Cornell
collaboration.  We present the convergence of the wave phase and
amplitude as the extrapolation order increases, and we also compare
data extrapolated using various extraction radii.  In
Sec.~\ref{sec:PhaseExtrapolation}, we propose a different
extrapolation method using the wave phase---similar to the method
introduced in Ref.~\cite{HannamEtAl2008}---to independently check our
results, again demonstrating the convergence properties of the method.
In Sec.~\ref{sec:Comparison}, we compare the extrapolated waveforms of
both methods at various extrapolation orders, showing that they agree
to well within the error estimates of the two methods.  A brief
discussion of the pitfalls and future of extrapolation is found in
Sec.~\ref{sec:Conclusions}.  Finally, we include a brief appendix on
techniques for filtering noisy data, which is particularly relevant
here because extrapolation amplifies noise.

%%%%%%%%%%%%%%%%%%%%%%%%%%%%%%%%%%%%%%%%%%%%%%%%%%%%%%%%%%%%%%%%%%%%%%
%%%%%%%%%%%%%%%%%%%%%%%%%%%%%%%%%%%%%%%%%%%%%%%%%%%%%%%%%%%%%%%%%%%%%%
\section{Extrapolation using approximate tortoise coordinates}
\label{sec:TortoiseExtrapolation} %

There are many types of data that can be extracted from a numerical
simulation of an isolated source of gravitational waves.  The two most
common methods of extracting gravitational waveforms involve using the
Newman--Penrose $\Psi_{4}$ quantity, or the metric perturbation $\h$
extracted using Regge--Wheeler--Zerilli techniques.  Even if we focus
on a particular type of waveform, the data can be extracted at a
series of points along the $z$ axis, for example, or decomposed into
multipole components and extracted on a series of spheres around the
source.  To simplify this introductory discussion of extrapolation, we
ignore the variety of particular types of waveform data.  Rather, we
generalize to some abstract quantity $f$, which encapsulates the
quantity to be extrapolated and behaves roughly as a radially outgoing
wave.

We assume that $f$ travels along outgoing null cones, which we
parametrize by a retarded time $\tr$.  Along each of these null cones,
we further assume that $f$ can be expressed as a convergent (or at
least asymptotic) series in $1/r$---where $r$ is some radial
coordinate---for all radii of interest.  That is, we assume
\begin{equation}
  \label{eq:FormOfExtrapolatedFunction}
  f(\tr, r) = \sum_{k=0}^{\infty}\, \frac{f_{(k)}(\tr)} {r^{k}}\ ,
\end{equation}
for some functions $f_{(k)}$.  The asymptotic behavior of $f$ is given
by the lowest nonzero $f_{(k)}$.\footnote{For example, if
  $f=r\Psi_{4}$, then $f_{(0)}$ gives the asymptotic behavior; if
  $f=\Psi_{4}$, then $f_{(1)}$ gives the asymptotic behavior.}

Given data for such an $f$ at a set of retarded times, and a set of
radii $\{r_{i}\}$, it is a simple matter to fit the data for each
value of $\tr$ to a polynomial in $1/r$.  That is, for each value of
$\tr$, we take the set of data $\left\{ f(\tr, r_{i}) \right\}$ and
fit it to a finite polynomial so that
\begin{equation}
  \label{eq:FittingPolynomial}
  f(\tr, r_{i}) \simeq \sum_{k=0}^{N}\, \frac{f_{(k)}(\tr)}
  {r^{k}_{i}}\ .
\end{equation}
Standard algorithms~\cite{PressEtAl2007} can be used to accomplish
this fitting; here we use the least-squares method.  Of course,
because we are truncating the series of
Eq.~\eqref{eq:FormOfExtrapolatedFunction} at $k=N$, some of the
effects from $k>N$ terms will appear at lower orders.  We will need to
choose $N$ appropriately, checking that the extrapolated quantity has
converged sufficiently with respect to this order.

\subsection{Radial parameter}
\label{sec:ChoiceOfR}
One subtlety to be considered is the choice of $r$ parameter to be
used in the extraction and fitting.  For numerical simulation of an
isolated system, one simple and obvious choice is the coordinate
radius $\rc$ used in the simulation.  Alternatively, if the data is
measured on some spheroidal surface, it is possible to define an areal
radius $\ra$ by measuring the area of the sphere along with $f$, and
setting $\ra \equiv \sqrt{\text{area}/4\pi}$.  Still other choices are
certainly possible.

One objective in choosing a particular $r$ parameter is to ensure the
physical relevance of the final extrapolated quantity.  If we try to
detect the wave, for example, we may want to think of the detector as
being located at some constant value of $r$.  Or, we may want $r$ to
asymptotically represent the luminosity distance.  These conditions
may be checked by inspecting the asymptotic behavior of the metric
components in the given coordinates.  For example, if the metric
components in a coordinate system including $r$ asymptotically
approach those of the standard Minkowski metric, it is not hard to see
that an inertial detector could follow a path of constant $r$
parameter.

Suppose we have two different parameters $r$ and $\tilde{r}$ which can
be related by a series expansion
\begin{equation}
  \label{eq:rRelation}
  r = \tilde{r}\, \left[ 1 + a/\tilde{r} + \ldots \right]\ .
\end{equation}
For the data presented in this paper, we can show that the coordinate
radius $\rc$ and areal radius $\ra$ are related in this way.
Introducing the expansion coefficients $\tilde{f}_{(k)}$, we can write
\begin{equation}
  \label{eq:FormOfExtrapolatedFunctionExpanded}
  f(\tr, r) = \sum_{k=0}^{\infty}\, \frac{f_{(k)}(\tr)} {r^{k}} =
  \sum_{k=0}^{\infty}\, \frac{\tilde{f}_{(k)}(\tr)} {\tilde{r}^{k}}\ .
\end{equation}
Inserting Eq.~\eqref{eq:rRelation} into this formula, Taylor
expanding, and equating terms of equal order $k$, shows that $f_{(0)}
= \tilde{f}_{(0)}$ and $f_{(1)} = \tilde{f}_{(1)}$.  Thus, if the
asymptotic behavior of $f$ is given by $f_{(0)}$ or $f_{(1)}$, the
final extrapolated data should not depend on whether $r$ or
$\tilde{r}$ is used.  On the other hand, in practice we truncate these
series at finite order.  This means that higher-order terms could
``pollute'' $f_{(0)}$ or $f_{(1)}$.  The second objective in choosing
an $r$ parameter, then, is to ensure fast convergence of the series in
Eq.~\eqref{eq:FittingPolynomial}.  If the extrapolated quantity does
not converge quickly as the order of the extrapolating polynomial $N$
is increased, it may be due to a poor choice of $r$ parameter.

The coordinate radius used in a simulation may be subject to large
gauge variations that are physically irrelevant, and hence are not
reflected in the wave's behavior.  That is, the wave may not fall off
nicely in inverse powers of that coordinate radius.  For the data
discussed later in this paper, we find that using the coordinate
radius of extraction spheres is indeed a poor choice, while using the
areal radius of those extraction spheres improves the convergence of
the extrapolation.

\subsection{Retarded-time parameter}
\label{sec:ChoiceOfRetardedTime} %

Similar considerations must be made for the choice of retarded-time
parameter $\tr$ to be used in extrapolation.  It may be possible to
evolve null geodesics in numerical simulations, and use these to
define the null curves on which data is to be extracted.  While this
is an interesting possibility that deserves investigation, we propose
two simpler methods here based on an approximate retarded time
constructed using the coordinates of the numerical simulation and the
phase of the waves measured in that coordinate system.

Again, we have two criteria for choosing a retarded-time parameter.
First is the physical suitability in the asymptotic limit.  For
example, we might want the asymptotic $\tr$ to be (up to an additive
term constant in time) the proper time along the path of a detector
located at constant $r$.  Again, checking the asymptotic behavior of
the metric components with respect to $\tr$ and $r$ should be a
sufficient test of the physical relevance of the parameters.  Second,
we wish to have rapid convergence of the extrapolation series using
the chosen parameter, which also needs to be checked.

As before, we can also show the equivalence of different choices for
the $\tr$ parameter.  Suppose we have two different approximations
$\tr$ and $\trs$ that can be related by a series expansion
\begin{equation}
  \label{eq:trRelation}
  \tr = \trs\, \left[ 1 + b/r + \ldots \right]\ .
\end{equation}
Using the new expansion coefficients $\accentset{\smile}{f}_{(k)}$, we
can write
\begin{equation}
  \label{eq:FormOfExtrapolatedFunctionExpandedForTR}
  f(\tr, r) = \sum_{k=0}^{\infty}\, \frac{f_{(k)}(\tr)} {r^{k}} =
  \sum_{k=0}^{\infty}\, \frac{\accentset{\smile}{f}_{(k)}(\trs)}
  {r^{k}}\ .
\end{equation}
Now, however, we need to assume that the functions $f_{(k)}$ can be
well-approximated by Taylor series.  If this is true, we can again
show that $f_{(0)} = \accentset{\smile}{f}_{(0)}$ or, \emph{if} we
have $f_{(0)}=\accentset{\smile}{f}_{(0)}=0$, that $f_{(1)} =
\accentset{\smile}{f}_{(1)}$.  The condition that $f$ be
well-approximated by a Taylor series is nontrivial, and can help to
inform the choice of $f$.  Similarly, the speed of convergence of the
extrapolation can help to inform the choice of a particular $\tr$
parameter.  While it has been shown \cite{DamourEtAl2008} that a
retarded-time parameter as simple as $\tr=T-R$ is sufficient for some
purposes, we find that convergence during and after merger is
drastically improved when using a somewhat more careful choice.

Since we will be considering radiation from an isolated compact
source, our basic model for $\tr$ comes from the Schwarzschild
spacetime; we assume that the system in question approaches this
spacetime at increasing distance.  In analogy with the
time-retardation effect on outgoing null rays in a Schwarzschild
spacetime~\cite{Chandrasekhar1992}, we define a ``tortoise
coordinate'' $\rt$ by:
\begin{equation}
  \label{eq:TortoiseCoordinate}
  \rt \equiv r + 2 \ADMMass \ln\left(\frac{r}{2\ADMMass} - 1 \right)\
  ,
\end{equation}
where $\ADMMass$ is the ADM mass of the initial data.\footnote{ Kocsis
  and Loeb~\cite{KocsisLoeb2007} pointed out that the propagation of a
  roughly spherical gravitational wave should be affected primarily by
  the amount of mass \emph{interior to} the wave.  Because the waves
  from a merging binary can carry off a significant fraction
  (typically a few percent) of the binary's mass, this suggests that
  we should allow the mass in this formula to vary in time, falling by
  perhaps a few percent over the duration of the waveform.  However,
  this is a small correction of a small correction; we have not found
  it necessary.  Perhaps with more refined methods, this additional
  correction would be relevant.} %
In standard Schwarzschild coordinates, the appropriate retarded time
would be given by $\tr = t - \rt$.  It is not hard to see that the
exterior derivative $\d \tr$ is null with respect to the Schwarzschild
metric.

Taking inspiration from this, we can attempt to account for certain
differences from a Schwarzschild background.  Let $T$ and $R$ denote
the simulation's coordinates, and suppose that we extract the metric
components $g^{TT}$, $g^{TR}$, and $g^{RR}$ from the simulation.  We
seek a $\tr(T,R)$ such that
\begin{equation}
  \label{eq:dTRetarded}
  \d\tr = \frac{\partial \tr}{\partial T}\, \d T +\frac{\partial
    \tr}{\partial R}\, \d R
\end{equation}
is null with respect to these metric components.  That is, we seek a
$\tr$ such that
\begin{multline}
  \label{eq:NullCondition}
  g^{TT}\, \left( \frac{\partial \tr}{\partial T} \right)^{2} + 2
  g^{TR}\, \left( \frac{\partial \tr}{\partial T} \right)\, \left(
    \frac{\partial \tr}{\partial R} \right) \\ + g^{RR}\, \left(
    \frac{\partial \tr}{\partial R} \right)^{2} = 0\ .
\end{multline}
We introduce the ansatz $\tr = t - \rt$, where $t$ is assumed to be a
slowly varying function of $R$,\footnote{More specifically, we need
  $\lvert \partial t/\partial R \rvert \ll \lvert \partial \rt
  / \partial R \rvert$.  This condition needs to be checked for all
  radii used, at all times in the simulation.  For the data presented
  below, we have checked this, and shown it to be a valid assumption,
  at the radii used for extrapolation.} %
and $\rt$ is given by Eq.~\eqref{eq:TortoiseCoordinate} with $R$ in
place of $r$ on the right side.  If we ignore $\partial t / \partial
R$ and insert our ansatz into Eq.~\eqref{eq:NullCondition}, we have
\begin{multline}
  \label{eq:NullConditionB}
  g^{TT}\, \left( \frac{\partial t}{\partial T} \right)^{2} - 2
  g^{TR}\, \left( \frac{\partial t}{\partial T} \right)\, \left(
    \frac{1}{1-2\ADMMass/R} \right) \\ + g^{RR}\, \left(
    \frac{1}{1-2\ADMMass/R} \right)^{2} = 0\ .
\end{multline}
We can solve this for $\partial t / \partial T$:
\begin{equation}
  \label{eq:RetardedTimeSolutionA}
  \frac{\partial t} {\partial T} = \frac{1}{1-2\ADMMass/R}\,
  \frac{g^{TR} \pm \sqrt{(g^{TR})^{2} - \, g^{TT}\, g^{RR}}} {g^{TT}}
  \ .
\end{equation}
Substituting the Schwarzschild metric components shows that we should
choose the negative sign in the numerator of the second factor.
Finally, we can integrate (numerically) to find
\begin{equation}
  \label{eq:FullRetardedTimeSolution}
  t = \int_{0}^{T}\, \frac{1}{g^{TT}}\, \frac{g^{TR} -
    \sqrt{(g^{TR})^{2} - g^{TT}\, g^{RR}}} {1-2\ADMMass/R}\, \d T'\ .
\end{equation}
Now, in the case where $g^{TR}$ is small compared to 1, we may wish to
ignore it, in which case we have
\begin{equation}
  \label{eq:RetardedTimeSolutionB}
  t = \int_{0}^{T}\, \frac{\sqrt{-g^{RR} / g^{TT}}} {1-2\ADMMass/R}\,
  \d T'\ .
\end{equation}
It is not hard to see that this correctly reduces to $t=T$ in the
Schwarzschild case.

For the data discussed later in this paper, we make further
assumptions that $g^{RR} = 1-2\ADMMass/R$, and that $R=\ra$.  That is,
we define the corrected time
\begin{subequations}
  \label{eq:DynamicLapseCorrection}
  \begin{gather}
    \tcorr \equiv \int_{0}^{T}\, \sqrt{\frac{-1/g^{TT}}{1 -
        2\ADMMass/\ra}} \, \d T' \intertext{and the retarded time} \tr
    \equiv \tcorr - \rt\ .
  \end{gather}
\end{subequations}
We find that this corrected time leads to a significant improvement
over the naive choice of $t(T)=T$, while no improvement results from
using Eq.~\eqref{eq:FullRetardedTimeSolution}.

\subsection{Application to a binary inspiral}
\label{sec:Application}
To begin the extrapolation procedure, we extract the (spin-weight
$s=-2$) $(l,m)=(2,2)$ component of $\Psi_{4}$ data on a set of spheres
at constant coordinate radius in the simulation.\footnote{See
  Ref.~\cite{ScheelEtAl2009} for details of the extraction procedure.
  We use $\Psi_{4}$ data here, rather than Regge--Wheeler--Zerilli
  data because the $\Psi_{4}$ data from this simulation is of higher
  quality; it appears that the RWZ data is more sensitive to changes
  in gauge conditions after the merger.  This problem is still under
  investigation.} %
In the black-hole binary simulations used here (the same as those
discussed in Refs.~\cite{BoyleEtAl2007b, Boyle2008, BoyleEtAl2008a,
  ScheelEtAl2009}), these spheres are located roughly\footnote{
  Explicitly, the extraction spheres are at radii $\rc/\Mirr = \{75,
  85, 100, 110, 120, \ldots, 190, 200, 210, 225\}$, though we find
  that the final result is not sensitive to the exact placement of the
  extraction spheres.} %
every $\Delta\rc \approx 10\IrrMass$ (where $\IrrMass$ is the sum of
the irreducible masses of the black holes in the initial data) from an
inner radius of $\rc=75\IrrMass$ to an outer radius of
$\rc=225\IrrMass$, where $\IrrMass$ denotes the total apparent-horizon
mass of the two holes at the beginning of the simulation.  This
extraction occurs at time steps of $\Delta \tc \approx 0.5\IrrMass$
throughout the simulation.  We also measure the areal radius, $\ra$,
of these spheres by integrating the induced area element over the
sphere to find the area, and defining $\ra \equiv
\sqrt{\text{area}/4\pi}$.  This typically differs from the coordinate
radius $\rc$ by roughly $\IrrMass/\rc$.  Because of gauge effects, the
areal radius of a coordinate sphere changes as a function of time, so
we measure this as a function of time.  Finally, we measure the
average value of $g^{TT}$ as a function of coordinate time on the
extraction spheres to correct for the dynamic lapse function.  The
areal radius and $g^{TT}$ are then used to compute the retarded time
$\tr$ defined in Eq.~\eqref{eq:DynamicLapseCorrection}.

The gravitational-wave data $\Psi_{4}$, the areal radius $\ra$, and
the lapse $N$ are all measured as functions of the code coordinates
$\tc$ and $\rc$.  We can use these to construct the retarded time
defined in Eq.~\eqref{eq:DynamicLapseCorrection}, using $\ra$ in place
of $r$.  This, then, will also be a function of the code coordinates.
The mapping between $(\tr,\ra)$ and $(\tc,\rc)$ is invertible, so we
can rewrite $\Psi_{4}$ as a function of $\tr$ and $\ra$.

As noted in Sec.~\ref{sec:ChoiceOfRetardedTime}, we need to assume
that the extrapolated functions are well approximated by Taylor
series.  Because the real and imaginary parts of $\Psi_{4}$ are
rapidly oscillating in the data presented here, we prefer to use the
same data in smoother form.  We define the complex amplitude $A$ and
phase $\phi$ of the wave:
\begin{equation}
  \label{eq:AmplitudeAndPhaseDefinition}
  % \label{eq:PsiFourAmplitudeAndPhaseDefinition}
  \ra\, \IrrMass\, \Psi_4 \equiv A\, \e^{\i \phi}\ ,
\end{equation}
where $A$ and $\phi$ are functions of $\tr$ and $\ra$.  Note that this
definition factors out the dominant $1/r$ behavior of the amplitude.
This equation defines the phase with an ambiguity of multiples of
$2\pi$.  In practice, we ensure that the phase is continuous as a
function of time by adding suitable multiples of $2\pi$.  The
continuous phase is easier to work with for practical reasons, and is
certainly much better approximated by a Taylor series, as required by
the argument surrounding
Eq.~\eqref{eq:FormOfExtrapolatedFunctionExpandedForTR}.

A slight complication arises in the relative phase offset between
successive radii.  Noise in the early parts of the waveform makes the
overall phase offset go through multiples of $2\pi$ essentially
randomly.  We choose some fairly noise-free (retarded) time and ensure
that phases corresponding to successive extraction spheres are matched
at that time, by simply adding multiples of $2\pi$ to the phase of the
entire waveform---that is, we add a multiple of $2\pi$ to the phase at
all times.

Extrapolation of the waveform, then, basically consists of finding the
asymptotic forms of these functions, $A$ and $\phi$ as functions of
time.  We apply the general technique discussed above to $A$ and
$\phi$.  Explicitly, we fit the data to polynomials in $1/\ra$ for
each value of retarded time:
\begin{subequations}
  \label{eq:ExtrapolationFormula}
  \begin{align}
    \label{eq:AxmplitudeExtrapolation}
    A(\tr,\ra) &\simeq \sum_{k=0}^N\,\frac{A_{(k)}(\tr)}{\ra^k}\ , \\
    \label{eq:PhaseExtrapolation}
    \phi(\tr,\ra) &\simeq \sum_{k=0}^N\,\frac{\phi_{(k)}(\tr)}{\ra^k}\
    .
  \end{align}
\end{subequations}
The asymptotic waveform is fully described by $A_{(0)}$ and
$\phi_{(0)}$.  When the order of the approximating polynomials is
important, we will denote by $A_{N}$ and $\phi_{N}$ the asymptotic
waveforms resulting from approximations using polynomials of order
$N$.

We show the results of these extrapolations in the figures below.
Figs.~\ref{fig:LapseCorrectionComparison_Corr_Amp}
through~\ref{fig:LapseCorrectionComparison_NoCorrMatched} show
convergence plots for extrapolations using orders $N=1$--$5$.  The
first two figures show the relative amplitude and phase difference
between successive orders of extrapolation, using the corrected time
of Eq.~\eqref{eq:DynamicLapseCorrection}.  Here, we define
\begin{subequations}
  \begin{gather}
    \label{eq:RelativeAmplitudeDifferenceDefinition}
    \frac{\delta A}{A} \equiv \frac{A_{N_{a}} - A_{N_{b}}} {A_{N_{b}}}
    \\ \intertext{and}
    \label{eq:PhaseDifferenceDefinition}
    \delta \phi \equiv \phi_{N_{a}} - \phi_{N_{b}}\ .
  \end{gather}
\end{subequations}
When comparing waveforms extrapolated by polynomials of different
orders, we use $N_{b}=N_{a}+1$.  Note that the broad trend is toward
convergence, though high-frequency noise is more evident as the order
increases, as we discuss further in the next subsection.  The peak
amplitude of the waves occurs at time $\tr/\Mirr \approx 3954$.  Note
that the scale of the horizontal axis changes just before this time to
better show the merger/ringdown portion.  We see that the
extrapolation is no longer convergent, with differences increasing
slightly as the order of the extrapolating polynomial is increased.
The oscillations we see in these convergence plots have a frequency
equal to the frequency of the waves themselves.  Their origin is not
clear, but may be due to numerics, gauge, or other effects that
violate our assumptions about the outgoing-wave nature of the data.
It is also possible that there are simply no higher-order effects to
be extrapolated, so low-order extrapolation suffices.

Figure~\ref{fig:LapseCorrectionComparison_NoCorrMatched} shows the
same data as in Fig.~\ref{fig:LapseCorrectionComparison_Corr}, except
that no correction is used for dynamic lapse.  That is, for this
figure (and only this figure), we use $\tr \equiv T - \rt$, where $T$
is simply the coordinate time.  This demonstrates the need for
improved time-retardation methods after merger.  Note that the
extrapolated data during the long inspiral is virtually unchanged
(note the different vertical axes).  After the merger---occurring at
roughly $\tr/\Mirr = 3954$---there is no convergence when no
correction is made for dynamic lapse.  It is precisely the merger and
ringdown segment during which extreme gauge changes are present in the
data used here~\cite{ScheelEtAl2009}.  On the other hand, the fair
convergence of the corrected waveforms indicates that it is possible
to successfully remove these gauge effects.

%%%%%%%%%%%%%%%%%%%%%%%%%%%%%%%%%%%%%%%%%%%%%%%%%%%%%%%%%%%%%%%%%%%%%%
\begin{figure}
  %LapseCorrectionComparison_Corr_Amp
  \includegraphics[width=\linewidth]{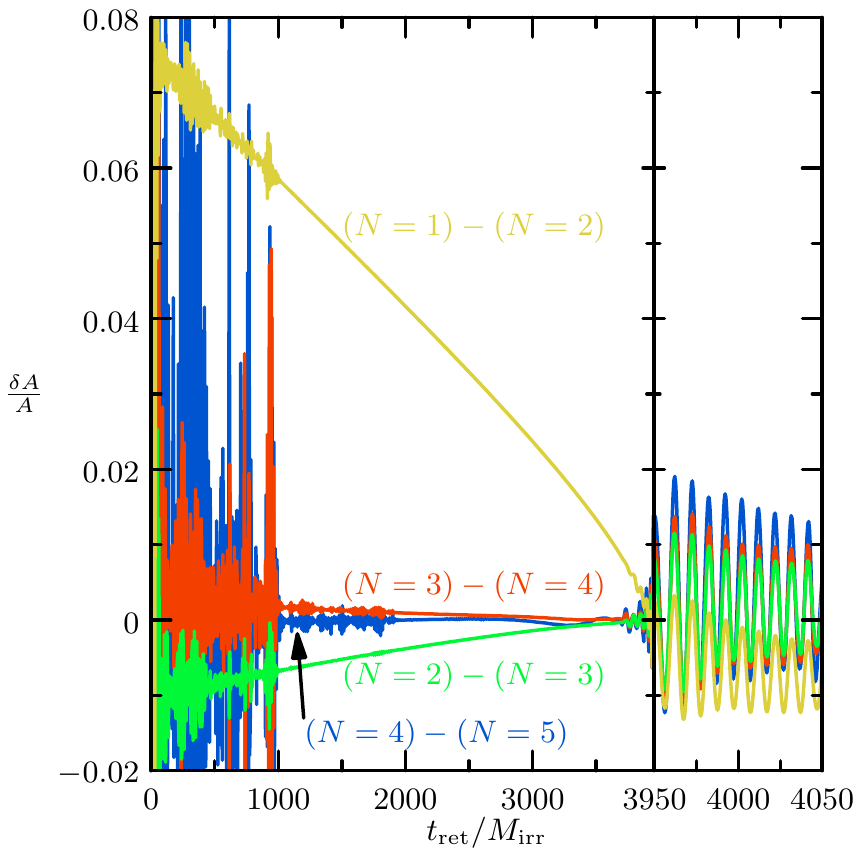}
  \caption{\CapName{Convergence of the amplitude of the extrapolated
      $\Psi_{4}$, with increasing order of the extrapolating
      polynomial, $N$} This figure shows the convergence of the
    relative amplitude of the extrapolated Newman--Penrose waveform,
    as the order $N$ of the extrapolating polynomial is increased.
    (See Eq.~\eqref{eq:ExtrapolationFormula}.)  That is, we subtract
    the amplitudes of the two waveforms, and normalize at each time by
    the amplitude of the second waveform.  We see that increasing the
    order tends to amplify the apparent noise during the early and
    late parts of the waveform.  Nonetheless, the broad
    (low-frequency) trend is towards convergence.  Note that the
    differences decrease as the system nears merger; this is a first
    indication that the extrapolated effects are due to near-field
    influences.  Also note that the horizontal axis changes in the
    right part of the figure, which shows the point of merger, and the
    ringdown portion of the waveform.  After the merger, the
    extrapolation is nonconvergent, though the differences grow slowly
    with the order of extrapolation.}
  \label{fig:LapseCorrectionComparison_Corr_Amp}
\end{figure}
%%%%%%%%%%%%%%%%%%%%%%%%%%%%%%%%%%%%%%%%%%%%%%%%%%%%%%%%%%%%%%%%%%%%%%

%%%%%%%%%%%%%%%%%%%%%%%%%%%%%%%%%%%%%%%%%%%%%%%%%%%%%%%%%%%%%%%%%%%%%%
\begin{figure}
  %LapseCorrectionComparison_Corr
  \includegraphics[width=\linewidth]{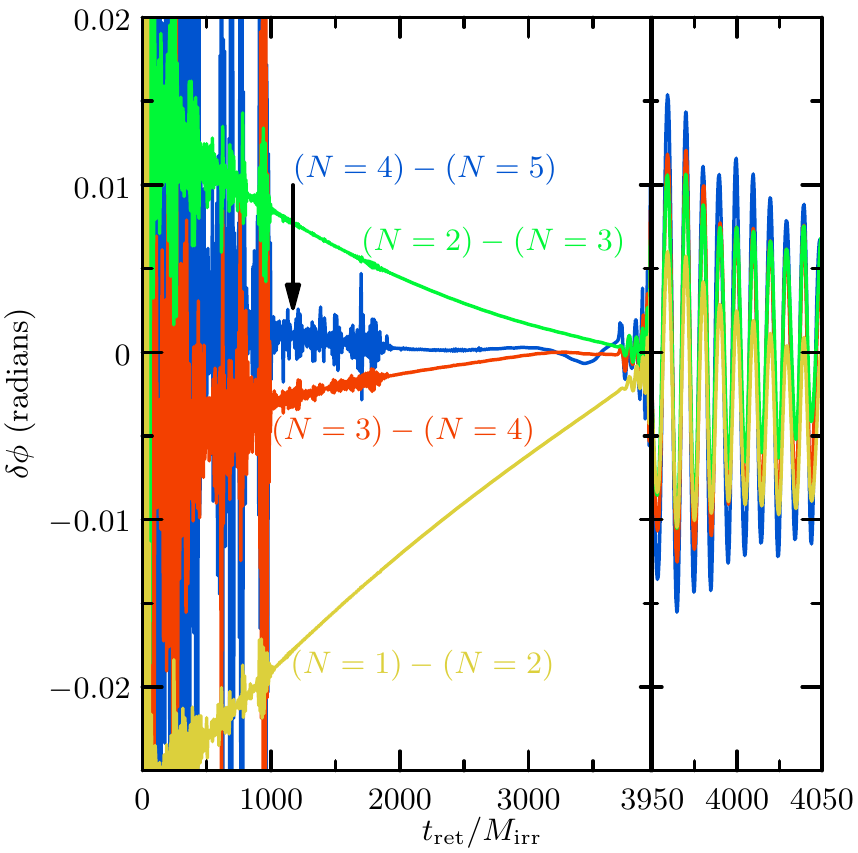}
  \caption{\CapName{Convergence of the phase of the extrapolated
      $\Psi_{4}$, with increasing order of the extrapolating
      polynomial, $N$} This figure is the same as
    Fig.~\ref{fig:LapseCorrectionComparison_Corr_Amp}, except that it
    shows the convergence of phase.  Again, increasing the
    extrapolation order tends to amplify the noise during the early
    and late parts of the waveform, though the broad (low-frequency)
    trend is towards convergence.  The horizontal-axis scale changes
    just before merger.}
  \label{fig:LapseCorrectionComparison_Corr}
\end{figure}
%%%%%%%%%%%%%%%%%%%%%%%%%%%%%%%%%%%%%%%%%%%%%%%%%%%%%%%%%%%%%%%%%%%%%%

%%%%%%%%%%%%%%%%%%%%%%%%%%%%%%%%%%%%%%%%%%%%%%%%%%%%%%%%%%%%%%%%%%%%%%
\begin{figure}
  %LapseCorrectionComparison_NoCorrMatched
  \includegraphics[width=\linewidth]{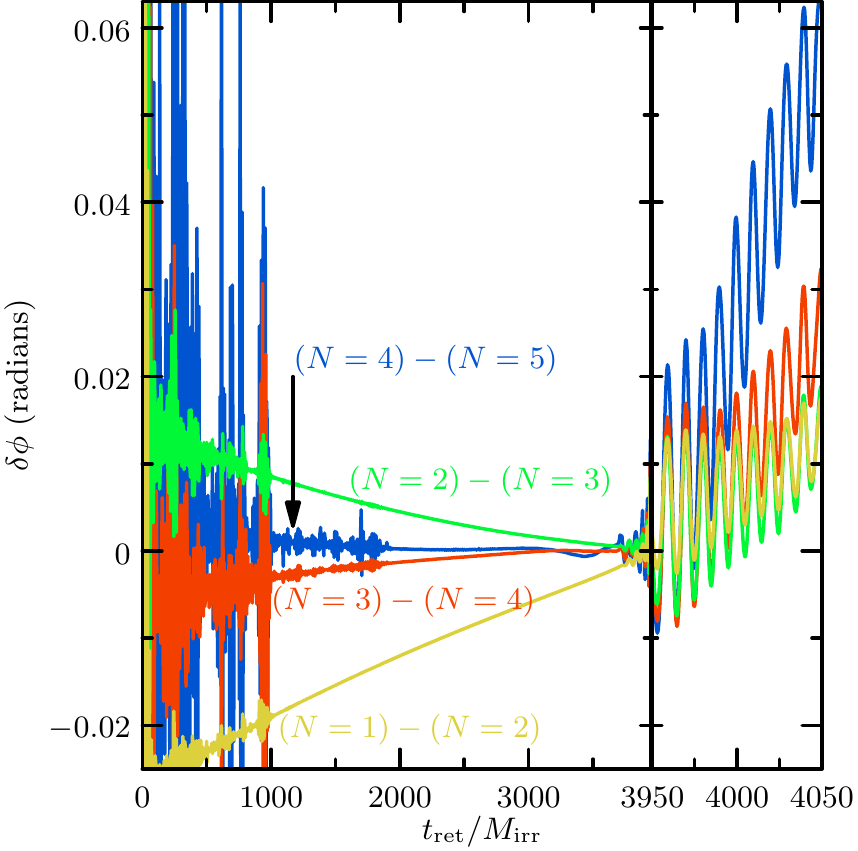}
  \caption{\CapName{Convergence of the phase of $\Psi_{4}$,
      extrapolated with no correction for the dynamic lapse} This
    figure is the same as
    Fig.~\ref{fig:LapseCorrectionComparison_Corr}, except that no
    correction is made to account for the dynamic lapse.  (See
    Eq.~\eqref{eq:DynamicLapseCorrection} and surrounding discussion.)
    Observe that the convergence is very poor after merger (at roughly
    $\tr/\Mirr = 3954$).  This corresponds to the time after which
    sharp features in the lapse are observed.  We conclude from this
    graph and comparison with the previous graph that the correction
    is crucial to convergence of $\Psi_{4}$ extrapolation through
    merger and ringdown.}
  \label{fig:LapseCorrectionComparison_NoCorrMatched}
\end{figure}
%%%%%%%%%%%%%%%%%%%%%%%%%%%%%%%%%%%%%%%%%%%%%%%%%%%%%%%%%%%%%%%%%%%%%%

\subsection{Choosing the order of extrapolation}
\label{sec:ExtrapolationOrder} %
Deciding on an appropriate order of extrapolation to be used for a
given purpose requires balancing competing effects.  As we see in
Fig.~\ref{fig:LapseCorrectionComparison_Corr_Amp}, for example, there
is evidently some benefit to be gained from using higher-order
extrapolation during the inspiral; there is clearly some convergence
during inspiral for each of the orders shown.  On the other hand,
higher-order methods amplify the apparent noise in the
waveform.\footnote{So-called ``junk radiation'' is a ubiquitous
  feature of initial data for current numerical simulations of binary
  black-hole systems.  It is clearly evident in simulations as
  large-amplitude, high-frequency waves that die out as the simulation
  progresses.  While it is astrophysically extraneous, it is
  nevertheless a correct result of evolution from the initial data.
  Better initial data would, presumably, decrease its magnitude.  This
  is the source of what looks like noise in the waveforms at early
  times.  It is less apparent in $\h$ data than in $\Psi_{4}$ data
  because $\Psi_{4}$ effectively amplifies high-frequency components,
  because of the relation $\Psi_{4} \approx -\ddot{h}$.} %
Moreover, late in the inspiral, and on into the merger and ringdown,
the effects being extrapolated may be present only at low orders;
increasing the extrapolation order would be useless as higher-order
terms would simply be fitting to noise.

The optimal order depends on the accuracy needed, and on the size of
effects that need to be eliminated from the data.  For some
applications, little accuracy is needed, so a low-order extrapolation
(or even no extrapolation) is preferable.\footnote{ We note that---as
  expected from investigations of near-field effects
  \cite{Teukolsky1982, Boyle2008, Boyle2009}---the second-order
  behavior of the amplitude greatly dominates its first-order behavior
  \cite{HannamEtAl2008}.  Thus, there is no improvement to the
  accuracy \emph{of the amplitude} when extrapolating with $N=1$; it
  would be better to simply use the data from the largest extraction
  radius.}  If high-frequency noise is not considered a problem, then
simple high-order extrapolation should suffice.  Of course, if both
high accuracy and low noise are required, data may easily be filtered,
mitigating the problem of noise amplification.  (See the appendix for
more discussion.)  There is some concern that this may introduce
subtle inaccuracies: filtering is more art than science, and it is
difficult to establish precise error bars for filtered data.

\subsection{Choosing extraction radii}
\label{sec:ExtractionRadii} %

Another decision needs to be made regarding the number and location of
extraction surfaces.  Choosing the number of surfaces is fairly easy,
because there is typically little cost in increasing the number of
extraction radii (especially relative to the cost of---say---running a
simulation).  The only restriction is that the number of data points
needs to be significantly larger than the order of the extrapolating
polynomial; more can hardly hurt.  More careful consideration needs to
be given to the \emph{location} of the extraction surfaces.

For the extrapolations shown in
Figs.~\ref{fig:LapseCorrectionComparison_Corr_Amp}
and~\ref{fig:LapseCorrectionComparison_Corr}, data was extracted on
spheres spaced by $10$ to $15\IrrMass$, from $R=75\IrrMass$ to
$R=225\IrrMass$.  The outer radius of $225\IrrMass$ was chosen simply
because this is the largest radius at which data exists throughout the
simulation; presumably, we always want the outermost radii at which
the data are resolved.  In choosing the inner radius, there are two
competing considerations.

%%%%%%%%%%%%%%%%%%%%%%%%%%%%%%%%%%%%%%%%%%%%%%%%%%%%%%%%%%%%%%%%%%%%%%
\begin{figure}
  %RadiiComparison
  \includegraphics[width=\linewidth]{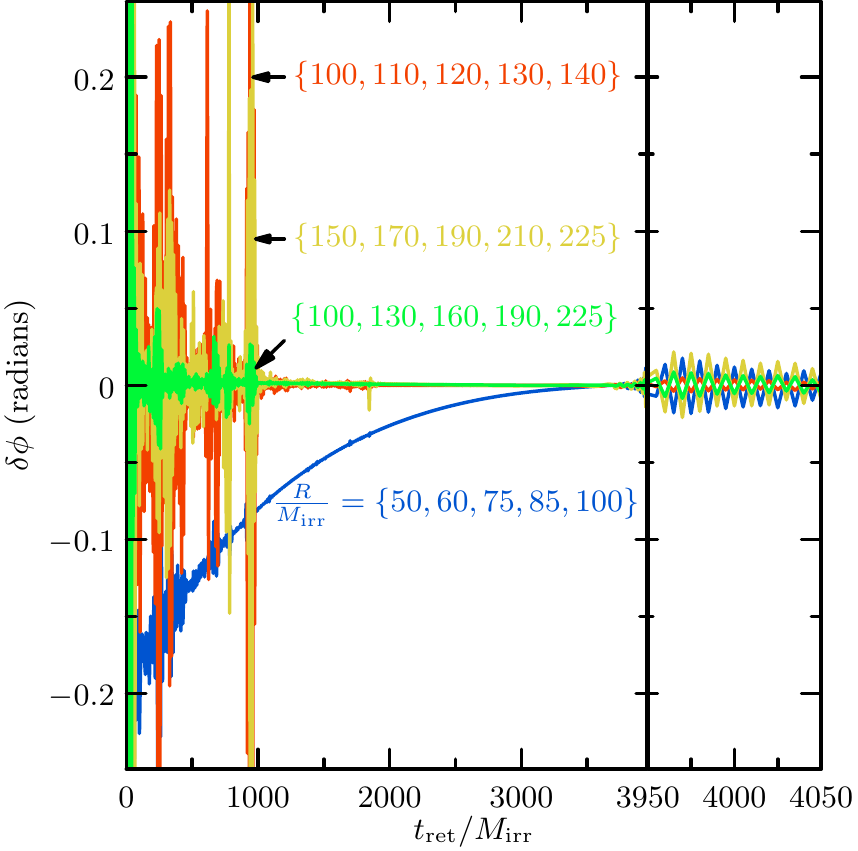}
  \caption{\CapName{Comparison of extrapolation of $\Psi_{4}$ using
      different sets of extraction radii} This figure compares the
    phase of waveforms extrapolated with various sets of radii.  All
    comparisons are with respect to the data set used elsewhere in
    this paper, which uses extraction radii $R/\IrrMass = \{75, 85,
    100, 110, 120, \ldots, 200, 210, 225\}$.  The order of the
    extrapolating polynomial is $N=3$ in all cases.}
  \label{fig:RadiiComparison}
\end{figure}
%%%%%%%%%%%%%%%%%%%%%%%%%%%%%%%%%%%%%%%%%%%%%%%%%%%%%%%%%%%%%%%%%%%%%%

On one hand, we want the largest spread possible between the inner and
outer extraction radii to stabilize the extrapolation.  A very rough
rule of thumb seems to be that the distance to be extrapolated should
be no greater than the distance covered by the data.  Because the
extrapolating polynomial is a function of $1/R$, the distance to be
extrapolated is $1/R_{\text{outer}} - 1/\infty = 1/R_{\text{outer}}$.
The distance covered by the data is $1/R_{\text{inner}} -
1/R_{\text{outer}}$, so if the rule of thumb is to be satisfied, the
inner extraction radius should be no more than half of the outer
extraction radius, $R_{\text{inner}} \lesssim R_{\text{outer}}/2$
(noting, of course, that this is a \emph{very} rough rule of thumb).

On the other hand, we would like the inner extraction radius to be as
far out as possible.  Extracting data near the violent center of the
simulation is a bad idea for many reasons.  Coordinate ambiguity,
tetrad errors, near-field effects---all are more severe near the
center of the simulation.  The larger these errors are, the more work
the extrapolation needs to do.  This effectively means that
higher-order extrapolation is needed if data are extracted at small
radii.  The exact inner radius needed for extrapolation depends on the
desired accuracy and, again, the portion of the simulation from which
the waveform is needed.

We can compare data extrapolated using different sets of radii.
Figure~\ref{fig:RadiiComparison} shows a variety, compared to the data
used elsewhere in this paper.  The extrapolation order is $N=3$ in all
cases.  Note that the waveforms labeled $R/\IrrMass = \{50, \ldots,
100\}$ and $R/\IrrMass = \{100, \ldots, 225\}$ both satisfy the rule
of thumb that the inner radius should be at most half of the outer
radius, while the other two waveforms do not; it appears that
violation of the rule of thumb leads to greater sensitivity to noise.
One waveform is extrapolated using only data from small radii,
$R/\IrrMass = \{50, \ldots, 100\}$.  It is clearly not converged, and
would require higher-order extrapolation if greater accuracy is
needed.  The source of the difference is presumably the near-field
effect~\cite{Boyle2008}, which is proportionally larger at small
radii.

Clearly, there is a nontrivial interplay between the radii used for
extraction and the order of extrapolation.  Indeed, because of the
time-dependence of the various elements of these choices, it may be
advisable to use different radii and orders of extrapolation for
different time portions of the waveform.  The different portions could
then be joined together using any of various
methods~\cite{AjithEtAl2008, BoyleEtAl2009a}.

%%%%%%%%%%%%%%%%%%%%%%%%%%%%%%%%%%%%%%%%%%%%%%%%%%%%%%%%%%%%%%%%%%%%%%
%%%%%%%%%%%%%%%%%%%%%%%%%%%%%%%%%%%%%%%%%%%%%%%%%%%%%%%%%%%%%%%%%%%%%%
\section{Extrapolation using the phase of the waveform}
\label{sec:PhaseExtrapolation}

While the tortoise-coordinate method just described attempts to
compensate for nontrivial gauge perturbations, it is possible that it
does not take account of all effects adequately. As an independent
check, we discuss what is essentially a second---very
different---formulation of the retarded-time parameter, similar to one
first introduced in Ref.~\cite{HannamEtAl2008}. If waves extrapolated
with the two different methods agree, then we can be reasonably
confident that unmodeled gauge effects are not diminishing the
accuracy of the final result.  As we will explain below, the method in
this section cannot be used naively with general data (\eg, data on
the equatorial plane).  In particular, we must assume that the data to
be extrapolated consists of a strictly monotonic phase.  It is,
however, frequently possible to employ a simple technique to make
purely real, oscillating data into complex data with strictly
monotonic phase, as we describe below.  The results of this technique
agree with those of the tortoise-coordinate extrapolation as we show
in Sec.~\ref{sec:Comparison}.

Instead of extrapolating the wave phase $\phi$ and amplitude $A$ as
functions of time and radius, we extrapolate the time $\tr$ and the
amplitude $A$ as functions of wave phase $\phi$ and radius $\ra$.  In
other words, we measure the amplitude and the arrival time to some
radius $\ra$ of a fixed phase point in the waveform.  This is the
origin of the requirement that the data to be extrapolated consist of
a strictly monotonic phase $\phi(\tr, \ra)$ (\ie, it must be
invertible).  For the data presented here, the presence of radiation
in the initial data---junk radiation---and numerical noise cause the
extracted waveforms to fail to satisfy this requirement at early
times.  In this case, the extrapolation is performed separately for
each invertible portion of the data.  That is, the data are divided
into invertible segments, each segment is extrapolated separately, and
the final products are joined together as a single waveform.

\subsection{Description of the method}
\label{sec:PhaseDescription} %
This extrapolation technique consists of extrapolating the retarded
time and the amplitude as functions of the wave phase $\phi$ and the
radius $\ra$.  In other words, when extrapolating the waveform, we are
estimating the amplitude and the arrival time of a fixed phase point
at infinity.  Here, we extract the same $\Psi_{4}$, $g^{TT}$, and
areal-radius data used in the previous section.  As in the previous
method, we first shift each waveform in time using $\tr = \tcorr -
\rt$, where $\tcorr$ is defined in
Eq.~\eqref{eq:DynamicLapseCorrection} and the basic tortoise
coordinate $\rt$ is defined in Eq.~\eqref{eq:TortoiseCoordinate} with
areal radius as the radial parameter.  The amplitude and wave phase
are again defined using Eq.~\eqref{eq:AmplitudeAndPhaseDefinition},
and the phase is made continuous as in Sec.~\ref{sec:Application}.
Thus, we begin with the same data, shifted as with the
tortoise-coordinate method.

Now, however, we change the method, in an attempt to allow for
unmodeled effects.  Instead of extrapolating $\phi(\tr, \ra)$ and
$A(\tr, \ra)$, as with the previous method, we invert these functions
to get $\tr(\phi,\ra)$ and $A(\phi,\ra)$ as functions of the wave
phase $\phi$.  In other words, we extrapolate the arrival time and the
amplitude of a signal to a coordinate radius $R$ for each wave phase
value.  This is done by fitting the retarded time $\tr$ and the
amplitude $A$ data to polynomials in $1/\ra$ for each value of the
wave phase:
\begin{subequations}
  \label{eq:ExtrapolationFormulaMethod2}
  \begin{align}
    \label{eq:AmplitudeExtrapolationMethod2}
    A(\ra, \phi) &\simeq \sum^N_{k=0}\frac{A_{(k)}(\phi)}{\ra^{k}}
    \ , \\
    \label{eq:PhaseExtrapolationMethod2}
    t(\ra, \phi) &\simeq \rt + \sum^N_{k=0}\frac{t_{(k)}(\phi)}{\ra^k}
    \ ,
  \end{align}
\end{subequations}
where the asymptotic waveform is fully described by $A_{(0)}(\phi)$
and $t_{(0)}(\phi)$.

With this data in hand, we can produce the asymptotic amplitude and
phase as functions of time by plotting curves in the \mbox{$t$--$A$}
and \mbox{$t$--$\phi$} planes parametrized by the phase.  In order to
be true, single-valued functions, we again need monotonicity of the
$t_{(0)}(\phi)$ data, which may be violated by extrapolation.  The
usable data can be obtained simply by removing data from times before
which this condition holds.

Choosing the extraction radii and extrapolation order for this method
follows the same rough recommendations described in
Secs.~\ref{sec:ExtrapolationOrder} and~\ref{sec:ExtractionRadii}.
Note also that the restriction that the data have an invertible phase
as a function of time is not insurmountable.  For example, data for
$\Psi_{4}$ in the equatorial plane is purely real, hence has a phase
that simply jumps from $0$ to $\pi$ discontinuously.  However, we can
define a new quantity
\begin{equation}
  w(t) \equiv \Psi_{4}(t) + \i \dot{\Psi}_{4}(t)\ .
\end{equation}
This is simply an auxiliary quantity used for the extrapolation, with
a smoothly varying, invertible phase.  The imaginary part is discarded
after extrapolation.

\subsection{Results}
\label{sec:PhaseResults} %

%%%%%%%%%%%%%%%%%%%%%%%%%%%%%%%%%%%%%%%%%%%%%%%%%%%%%%%%%%%%%%%%%%%%%%
\begin{figure}
  %AbdulConvergence_Amp
  \includegraphics[width=\linewidth]{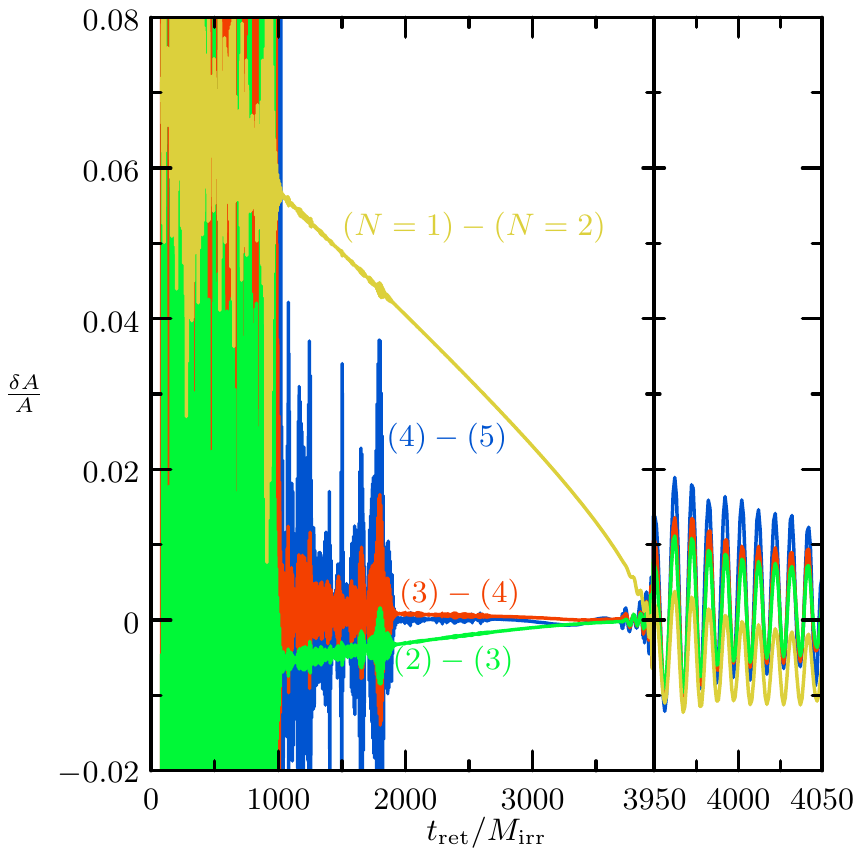}
  \caption{\CapName{Convergence of the amplitude of $\Psi_{4}$
      extrapolated using the wave phase, with increasing order $N$ of
      the extrapolating polynomial} This figure shows the convergence
    of the relative amplitude of the extrapolated Newman--Penrose
    waveform extrapolated using the wave phase, as the order $N$ of
    the extrapolating polynomial is increased.  (See
    Eq.~\eqref{eq:ExtrapolationFormulaMethod2}.)  Increasing the
    extrapolation order tends to amplify the apparent noise during the
    early and late parts of the waveform, but it improves convergence.
    The vertical axis at $\tr/\Mirr\approx 3950$ denotes merger.}
  \label{fig:AmplitudeComparisonMethod2}
\end{figure}
%%%%%%%%%%%%%%%%%%%%%%%%%%%%%%%%%%%%%%%%%%%%%%%%%%%%%%%%%%%%%%%%%%%%%%

\begin{figure}
  %AbdulConvergence_Phi
  \includegraphics[width=\linewidth]{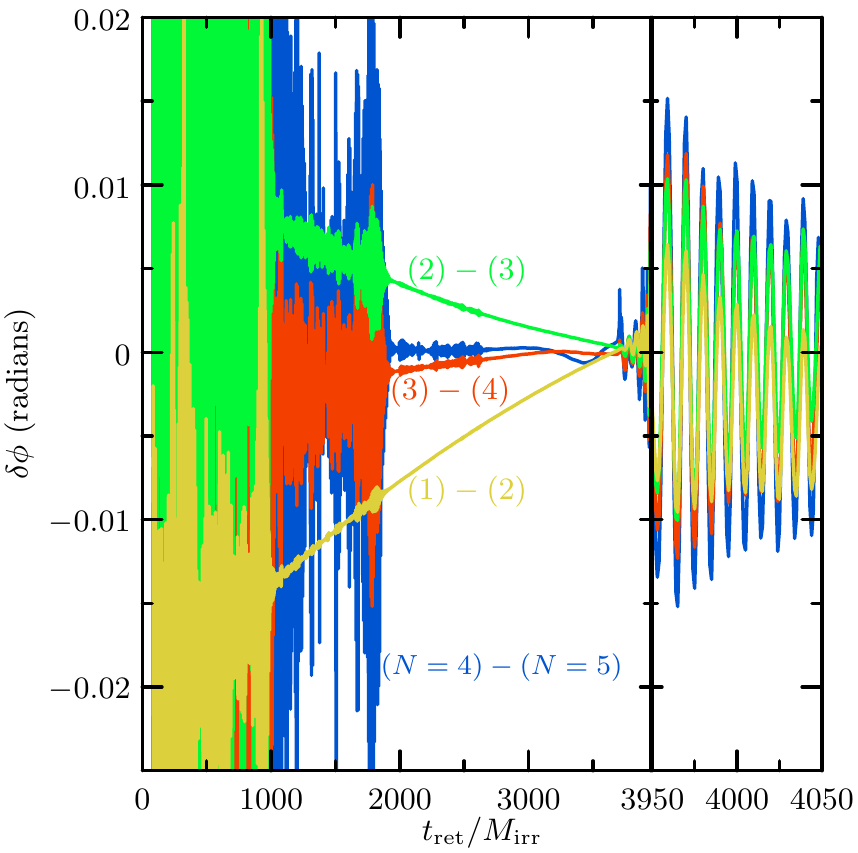}
  \caption{\CapName{Convergence of the phase of $\Psi_{4}$ as a
      function of time extrapolated using the wave phase, with
      increasing order $N$ of the extrapolating polynomial} Again,
    increasing the extrapolation order tends to amplify the apparent
    noise during the early and late parts of the waveform, though
    convergence is improved significantly.}
  \label{fig:PhaseComparisonMethod2}
\end{figure}

In Figs.~\ref{fig:AmplitudeComparisonMethod2}
and~\ref{fig:PhaseComparisonMethod2} we plot the convergence of the
relative amplitude and phase of the extrapolated $(l,m) = (2,2)$ mode
of the $\Psi_4$ waveform for extrapolation orders $N=1, \ldots, 5$.  A
common feature of both plots is that during the inspiral, the higher
the extrapolation order, the better the convergence.  However, the
noise is amplified significantly for large orders of extrapolation.
This method of extrapolation amplifies high-frequency noise
significantly, compared to the tortoise-coordinate method.

%%%%%%%%%%%%%%%%%%%%%%%%%%%%%%%%%%%%%%%%%%%%%%%%%%%%%%%%%%%%%%%%%%%%%%
%%%%%%%%%%%%%%%%%%%%%%%%%%%%%%%%%%%%%%%%%%%%%%%%%%%%%%%%%%%%%%%%%%%%%%

In the inspiral portion, we have a decreasing error in the
extrapolation of the phase and the amplitude as the wavelength of the
gravitational waves decreases.  In the merger/ringdown portion, a more
careful choice of the radii and order of extrapolation needs to be
made.  Since near-field effects are less significant in the data
extracted at larger radii, extrapolation at low order ($N=2,3$) seems
sufficient.  Data extrapolated at large order ($N=4,5$) has a larger
error in the phase and amplitude after merger than data extrapolated
at order $N=2$ or $3$.  Moreover, the outermost extraction radius
could be reduced, say to $R_{\text{outer}} / \Mirr = 165$ instead of
$R_{\text{outer}} / \Mirr = 225$, without having large extrapolation
error at late times.  Using the radius range $R/\Mirr={75,\ldots,160}$
instead of the range $R/ \Mirr = {75,\ldots,225}$ would leave the
extrapolation error during the merger/ringdown almost unchanged, while
the extrapolation error during the inspiral would increase by about
70\%.

We note that this method allows easy extrapolation of various portions
of the waveform using different extraction radii and orders since---by
construction---the wave phase is an independent variable. For example,
solve for the phase value of the merger $\phi_{\text{merger}}$
(defined as the phase at which the amplitude is a maximum), then use
the radius range $R / \Mirr = {75,\ldots,225}$ for all phase values
less than $\phi_{\text{merger}}$ and the range
$R/\Mirr={75,\ldots,160}$ for all larger phase values.

This method has been tested also using the coordinate radius $R$ and
the naive time coordinate $T$, in place of areal radius and corrected
time.  We found results similar to those discussed in
Sec.~\ref{sec:TortoiseExtrapolation}.  Using the new time coordinate
$\tcorr$ instead of the naive time coordinate $T$ improved the
extrapolation during the merger/ringdown, as found in
Sec.~\ref{sec:TortoiseExtrapolation}.

As with the previous extrapolation method, increasing the
extrapolation order gives a faster convergence rate of waveform phase
and amplitude, but it amplifies noise in the extrapolated waveform. To
improve convergence without increasing the noise, we need a good
filtering technique for the inspiral data. The junk-radiation noise
decreases significantly as a function of time, disappearing several
orbits before merger.  However, this noise could be reduced by using
more extraction radii in the extrapolation process, or by running the
data through a low-pass filter.  See the appendix for further
discussion of filtering.

%%%%%%%%%%%%%%%%%%%%%%%%%%%%%%%%%%%%%%%%%%%%%%%%%%%%%%%%%%%%%%%%%%%%%%
%%%%%%%%%%%%%%%%%%%%%%%%%%%%%%%%%%%%%%%%%%%%%%%%%%%%%%%%%%%%%%%%%%%%%%
\section{Comparison of the two methods}
\label{sec:Comparison}

Both methods showed good convergence of the amplitude and the phase of
the waveform as $N$ increased in the inspiral portion.  (See
Figs.~\ref{fig:LapseCorrectionComparison_Corr_Amp}
and~\ref{fig:AmplitudeComparisonMethod2} for the amplitude, and
Figs.~\ref{fig:LapseCorrectionComparison_Corr}
and~\ref{fig:PhaseComparisonMethod2} for the phase.) The wave-phase
extrapolation method was more sensitive to noise.  In the
merger/ringdown portion, both methods have similar convergence as $N$
increases, especially when the correction is taken to account for the
dynamic lapse.  The use of the time parameter $\tcorr$ improved the
agreement between the methods significantly in the merger/ringdown
portion for all extrapolation orders.  Extrapolating at order $N=2$ or
$3$ seems the best choice as the noise and phase differences are
smallest for these values.

\begin{figure}
  %MethodComparison_Amp
  \includegraphics[width=\linewidth]{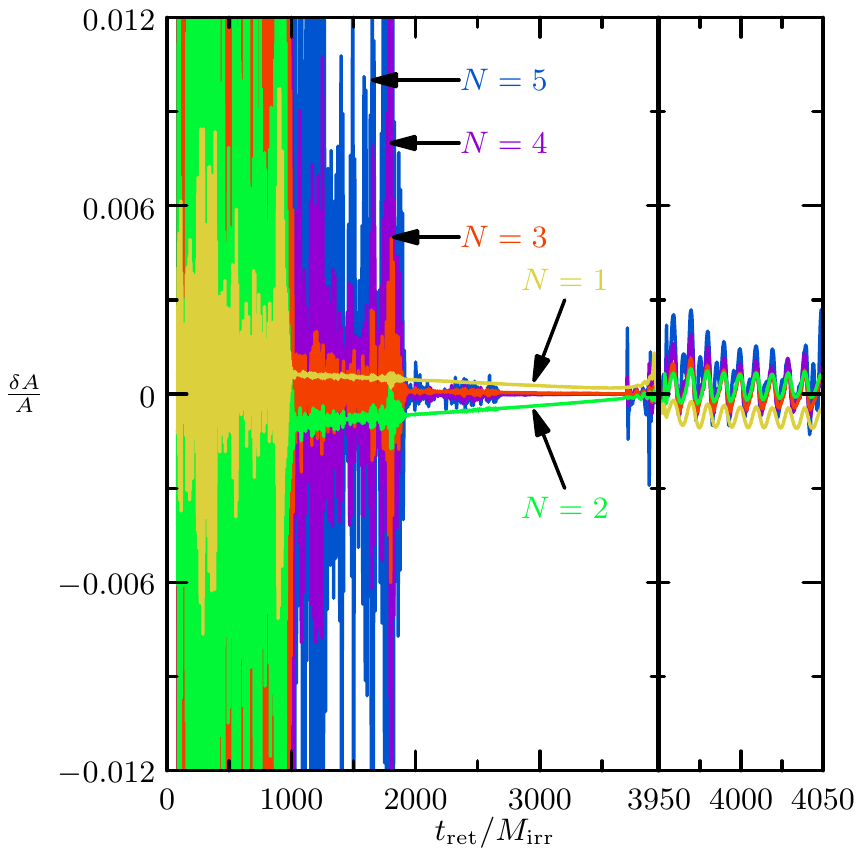}
  \caption{\CapName{Relative difference in the amplitude of the two
      extrapolation methods as we increase the order of extrapolation}
    The best agreement between both methods is at low orders of
    extrapolation, for which the relative difference in the amplitude
    is less than 0.1\% during most of the evolution.}
  \label{fig:MroueBoyleRelativeAmplitudeDifferenceByOrder}
\end{figure}

\begin{figure}
  %MethodComparison_Phi
  \includegraphics[width=\linewidth]{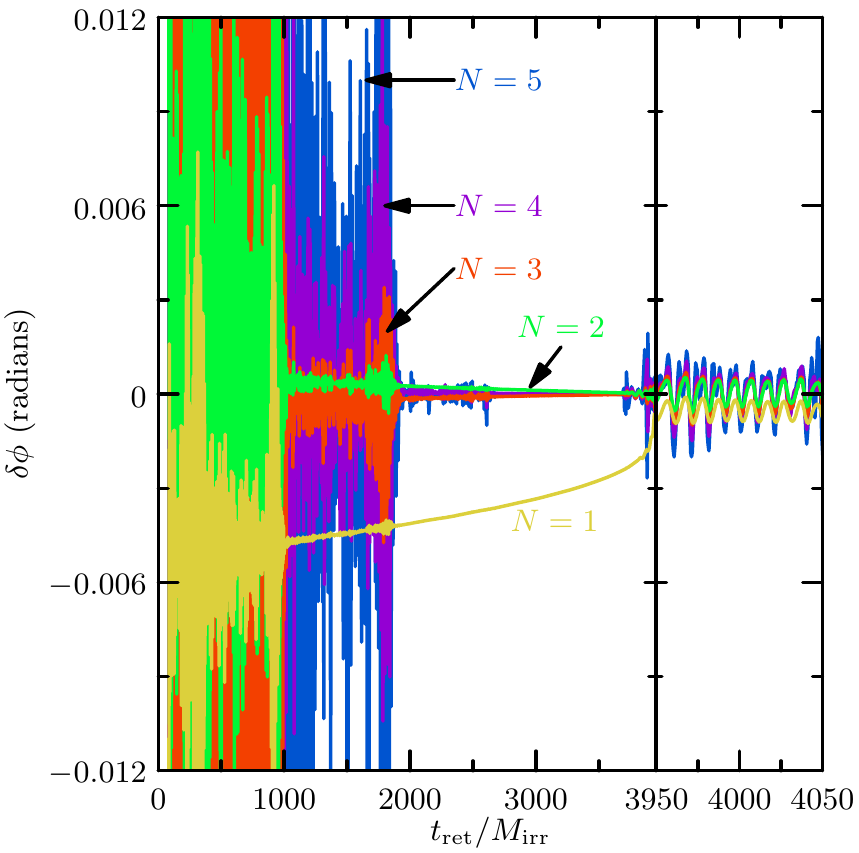}
  \caption{\CapName{Phase difference of the two extrapolation methods
      as we increase the order of extrapolation} This figure shows the
    phase difference between waveforms extrapolated using each of the
    two methods. The best agreement between the methods is at orders
    $N=2$ and $3$.  The relative difference in the phase is less than
    0.02 radians during most of the evolution.}
  \label{fig:MroueBoylePhaseDifferenceByOrder}
\end{figure}

In Fig.~\ref{fig:MroueBoyleRelativeAmplitudeDifferenceByOrder}, we
show the relative amplitude difference between data extrapolated at
various orders ($N=1,\ldots,5$).  There is no additional time or phase
offset used in the comparison.  Ignoring high-frequency components,
the difference in the relative amplitude is always less than 0.3\% for
different extrapolation orders.  Even including high-frequency
components, the differences between the two methods are always smaller
than the error in each method, as judged by convergence plots.  In
Fig.~\ref{fig:MroueBoylePhaseDifferenceByOrder}, we show the
\emph{phase} difference between the data extrapolated using both
methods.  As in the relative amplitude-difference plots, the best
agreement is achieved during the inspiral portion.  Ignoring
high-frequency components, the difference is less than 0.02 radians
for all orders.  In the merger/ringdown portion, the best agreement
between extrapolated waveforms is at order $N=2$ or $3$ where the
phase difference is less than 0.01 radians.

%%%%%%%%%%%%%%%%%%%%%%%%%%%%%%%%%%%%%%%%%%%%%%%%%%%%%%%%%%%%%%%%%%%%%%
%%%%%%%%%%%%%%%%%%%%%%%%%%%%%%%%%%%%%%%%%%%%%%%%%%%%%%%%%%%%%%%%%%%%%%

%%%%%%%%%%%%%%%%%%%%%%%%%%%%%%%%%%%%%%%%%%%%%%%%%%%%%%%%%%%%%%%%%%%%%%
%%%%%%%%%%%%%%%%%%%%%%%%%%%%%%%%%%%%%%%%%%%%%%%%%%%%%%%%%%%%%%%%%%%%%%

%%%%%%%%%%%%%%%%%%%%%%%%%%%%%%%%%%%%%%%%%%%%%%%%%%%%%%%%%%%%%%%%%%%%%%
%%%%%%%%%%%%%%%%%%%%%%%%%%%%%%%%%%%%%%%%%%%%%%%%%%%%%%%%%%%%%%%%%%%%%%
\section{Conclusions}
\label{sec:Conclusions}

We have demonstrated two simple techniques for extrapolating
gravitational-wave data from numerical-relativity simulations.  We
took certain basic gauge information into account to improve
convergence of the extrapolation during times of particularly dynamic
gauge, and showed that the two methods agree to within rough error
estimates.  We have determined that the first method presented here is
less sensitive to noise, and more immediately applies to arbitrary
wavelike data; this method has become the basic standard in use by the
Caltech--Cornell collaboration.  In both cases, there were problems
with convergence after merger.  The source of these problems is still
unclear, but will be a topic for further investigation.

As with any type of extrapolation, a note of caution is in order.  It
is entirely possible that the ``true'' function being extrapolated
bears little resemblance to the approximating function we choose,
outside of the domain on which we have data.  We may, however, have
reason to believe that the true function takes a certain form.  If the
data in question are generated by a homogeneous wave equation, for
instance, we know that well-behaved solutions fall off in powers of
$1/r$.  In any case, there is a certain element of faith that
extrapolation is a reasonable thing to do.  While that faith may be
misplaced, there are methods of checking whether or not it is:
goodness-of-fit statistics, error estimates, and convergence tests.
To be of greatest use, goodness-of-fit statistics and error estimates
for the output waveform require error estimates for the input
waveforms.  We leave this for future work.

We still do not know the correct answers to the questions numerical
relativity considers.  We have no analytic solutions to deliver the
waveform that Einstein's equations---solved perfectly---predict would
come from a black-hole binary merger; or the precise amount of energy
radiated from any given binary; or the exact kick or spin of the final
black hole.  Without being able to compare numerical relativity to
exact solutions, we may be leaving large systematic errors hidden in
plain view.  To eliminate them, we need to use multiple, independent
methods for our calculations.  For example, we might extract $\Psi_4$
directly by calculating the Riemann tensor and contracting
appropriately with our naive coordinate tetrad, and extract the metric
perturbation using the formalism of Regge--Wheeler--Zerilli and
Moncrief.  By differentiating the latter result twice and comparing to
$\Psi_4$, we could verify that details of the extraction methods are
not producing systematic errors.  (Just such a comparison was done in
Ref.~\cite{BuonannoEtAl2009a} for waveforms extrapolated using the
technique in this paper.)  Nonetheless, it is possible that
infrastructure used to find both could be leading to errors.

In the same way, simulations need to be performed using different
gauge conditions, numerical techniques, code infrastructures, boundary
conditions, and even different extrapolation methods.  Only when
multiple schemes arrive at the same result can we be truly confident
in any of them.  But to arrive at the same result, the waveforms from
each scheme need to be processed as carefully as possible.  We have
shown that extrapolation is crucial for highly accurate gravitational
waveforms, and for optimized detection of mergers in detector data.

%%%%%%%%%%%%%%%%%%%%%%%%%%%%%%%%%%%%%%%%%%%%%%%%%%%%%%%%%%%%%%%%%%%%%%
%%%%%%%%%%%%%%%%%%%%%%%%%%%%%%%%%%%%%%%%%%%%%%%%%%%%%%%%%%%%%%%%%%%%%%
%%%%%%%%%%%%%%%%%%%%%%%%%%%%%%%%%%%%%%%%%%%%%%%%%%%%%%%%%%%%%%%%%%%%%%
\begin{acknowledgments}
  We thank Emanuele Berti, Duncan Brown, Luisa Buchman, Alessandra
  Buonanno, Yanbei Chen, \'{E}anna Flanagan, Mark Hannam, Sascha Husa,
  Luis Lehner, Geoffrey Lovelace, Andrea Nerozzi, Rob Owen, Larne
  Pekowsky, Harald Pfeiffer, Oliver Rinne, Uli Sperhake, B\'{e}la
  Szil\'{a}gyi, Kip Thorne, Manuel Tiglio, and Alan Weinstein for
  helpful discussions.  We especially thank Larry Kidder, Lee
  Lindblom, Mark Scheel, and Saul Teukolsky for careful readings of
  this paper in various draft forms and helpful comments.  This
  research has been supported in part by a grant from the Sherman
  Fairchild Foundation to Caltech and Cornell; by NSF Grants
  PHY-0652952, PHY-0652929 and DMS-0553677 and NASA Grant NNX09AF96G
  to Cornell; and by NSF grants PHY-0601459, PHY-0652995, DMS-0553302
  and NASA grant NNG05GG52G to Caltech.
\end{acknowledgments}

%%%%%%%%%%%%%%%%%%%%%%%%%%%%%%%%%%%%%%%%%%%%%%%%%%%%%%%%%%%%%%%%%%%%%%
%%%%%%%%%%%%%%%%%%%%%%%%%%%%%%%%%%%%%%%%%%%%%%%%%%%%%%%%%%%%%%%%%%%%%%
\appendix* % Use \appendix* if there is just one appendix

\section{Filtering}

%%%%%%%%%%%%%%%%%%%%%%%%%%%%%%%%%%%%%%%%%%%%%%%%%%%%%%%%%%%%%%%%%%%%%%
\begin{figure}
  %FilteredDifference
  \includegraphics[width=\linewidth]{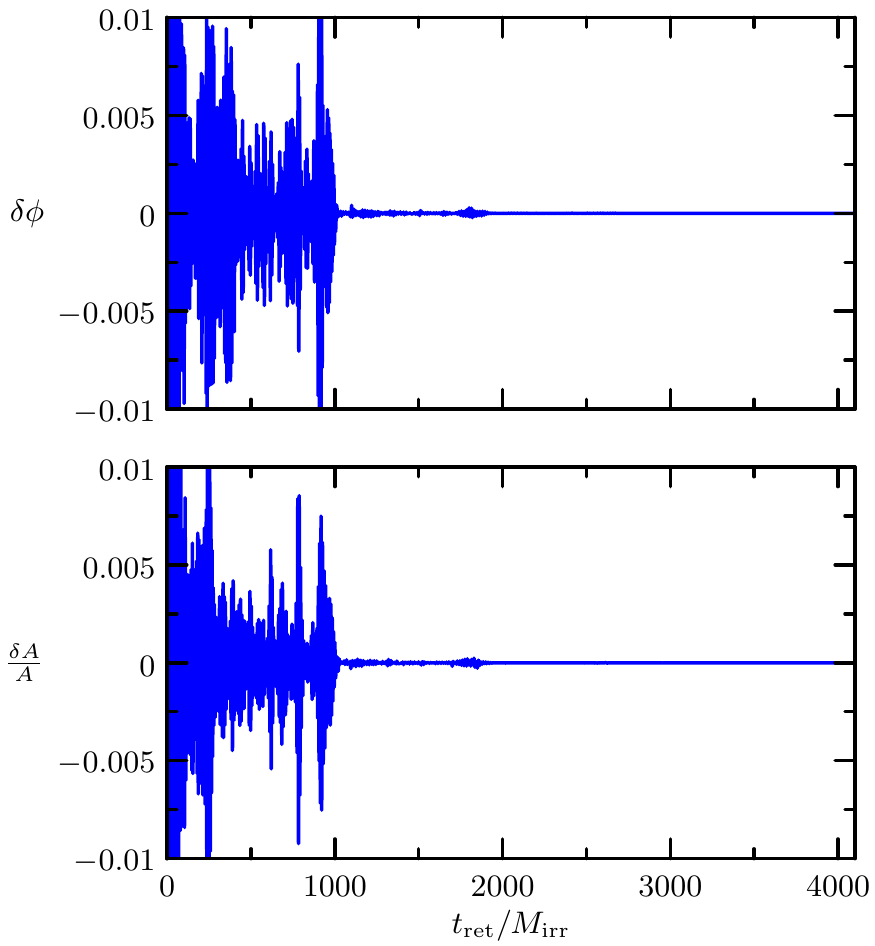}
  \caption{\CapName{Difference between the filtered and unfiltered
      amplitude and phase of the waveform with third-order
      extrapolation} The upper panel shows the relative amplitude
    difference between the filtered and unfiltered waveforms; the
    lower panel shows the phase difference.}
  \label{fig:Filtering}
\end{figure}
%%%%%%%%%%%%%%%%%%%%%%%%%%%%%%%%%%%%%%%%%%%%%%%%%%%%%%%%%%%%%%%%%%%%%%

Extrapolating waveforms containing poorly resolved high-frequency
components amplifies the magnitude of the noise in the signal at
infinity.  One possible solution to the problem is to filter out the
junk radiation from the gravitational waveform. This is possible when
the noise has higher frequency than the physical data of interest.
The \software{Matlab} function \command{filtfilt}, using a low-pass
Butterworth filter with cutoff frequency between the noise frequency
and the highest gravitational-wave frequency, is satisfactory for many
uses when applied to the early parts of the data.  This filtering may
be applied to either the complex data, or to its amplitude and
phase---the latter allowing for a lower cutoff frequency.  There is
also a marginal benefit to be gained when the input data are filtered
before extrapolation, though filtering of the final result is also
necessary.  It is also possible to fit a low-order polynomial to the
data, filter the residual, and add the filtered data back to the fit;
this removes very low-frequency components, reducing the impact of
Gibbs phenomena.

For the data presented here, we use a sixth-order Butterworth filter
with a physical cutoff frequency of $\omega_{\text{cutoff}} =
0.075/\Mirr$,\footnote{Note that \software{Matlab} expects the input
  frequency as a fraction of the data's Nyquist frequency.} which is
roughly eight times the maximum frequency of the physical waveforms
expected in the filtered region.  The filter is applied individually
(using the \command{filtfilt} function) to the amplitude and phase
data, in turn.  Because of remaining Gibbs phenomena at late times, we
use unfiltered data after $\tr/\Mirr = 3000$.

One basic diagnostic of the filtering process is to simply look at the
difference between filtered and unfiltered data.  If there are
low-frequency components in these curves, we know the cutoff frequency
needs to be raised.  In Fig.~\ref{fig:Filtering}, we show the
difference in relative amplitude (upper panel), and phase (lower
panel).  Because there is no difference between the filtered and
unfiltered waveforms on the timescale of the physical gravitational
waves ($\gtrsim 100\,\Mirr$), we conclude that the filter's cutoff
frequency is high enough to retain the physical information.

\begin{figure}
  %FilteredConvergencePlot
  \includegraphics[width=\linewidth]{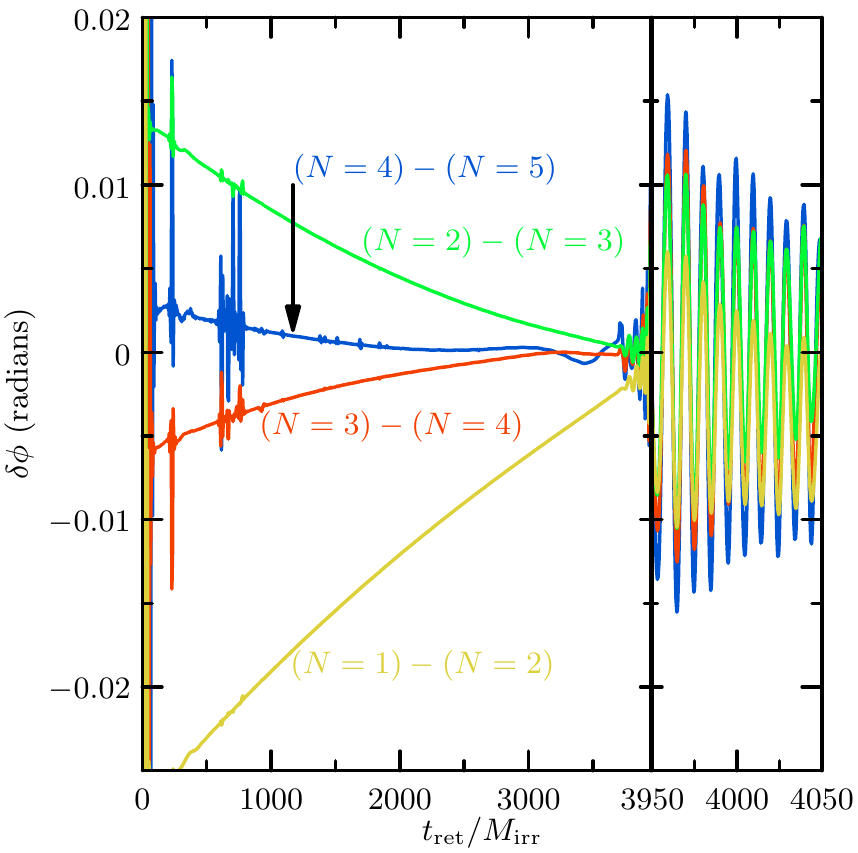}
  \caption{\CapName{The filtered version of
      Fig.~\ref{fig:LapseCorrectionComparison_Corr}} We filtered the
    extrapolated waveforms and redid
    Fig.~\ref{fig:LapseCorrectionComparison_Corr}, which shows the
    phase difference between waveforms extrapolated at various orders.
    This plot shows much smaller high-frequency components at early
    times.}
  \label{fig:FilteredConvergencePlot}
\end{figure}

On the other hand, to check that the filter's cutoff frequency is low
enough to achieve its purpose, we look at data which previously showed
the undesirable high-frequency characteristics.  In
Fig.~\ref{fig:FilteredConvergencePlot}, we show the same data as in
Fig.~\ref{fig:LapseCorrectionComparison_Corr}, when the data are
filtered before subtraction.  The size of the noise at early times is
greatly reduced.  There are still significant high-frequency features
in the plot, though they are much smaller than in the unfiltered data.
Presumably these features are simply so large in the input data that
even with the large suppression from the filter, they are still
noticeable.  It may be possible to remove them by further decreasing
the filter's cutoff frequency, though this would require better
handling of Gibb's phenomena from the beginning and end of the wave.
We find the present filter sufficient for the demonstration purposes
of this appendix.

%%%%%%%%%%%%%%%%%%%%%%%%%%%%%%%%%%%%%%%%%%%%%%%%%%%%%%%%%%%%%%%%%%%%%%
%%%%%%%%%%%%%%%%%%%%%%%%%%%%%%%%%%%%%%%%%%%%%%%%%%%%%%%%%%%%%%%%%%%%%%

%%%%%%%%%%%%%%%%%%%%%%%%%%%%%%%%%%%%%%%%%%%%%%%%%%%%%%%%%%%%%%%%%%%%%%
%%%%%%%%%%%%%%%%%%%%%%%%%%%%%%%%%%%%%%%%%%%%%%%%%%%%%%%%%%%%%%%%%%%%%%
\bibliography{References}

%%%%%%%%%%%%%%

\end{document}